\documentclass[12pt]{article}
\usepackage[utf8]{inputenc}
\usepackage[english, russian]{babel}
\usepackage{amsmath}
\usepackage{amsfonts}
\usepackage{amssymb}

\def\lb{\label}
\def\be{\begin{equation}}
\def\ee{\end{equation}}
\def\qed{\rule{5pt}{5pt}}

\newtheorem{proposition}{Proposition}

\begin{document}

\vspace{3cm}

	\begin{center}
		{\LARGE {$D$-dimensional
 spin projection operators for
  arbitrary type of symmetry
 via Brauer algebra idempotents
}}

 \vspace{2cm}

\large \sf
A.P.~Isaev$^{a,b,}$\footnote{\sf e-mail: isaevap@theor.jinr.ru},
M.A.~Podoinitsyn$^{a,}$\footnote{\sf e-mail: mikhailpodoinicin@gmail.com} \\

\vspace{1cm}
\item[$^a$]
{\it Bogoliubov Laboratory of Theoretical Physics, JINR, Dubna 141980, Russia}


 \vskip 0.5cm
 \ $^b${\it St.Petersburg Department of Steklov Mathematical
 Institute of RAS, \\ Fontanka 27, 191023 St. Petersburg, Russia}

\end{center}

\vspace{2cm}
\begin{otherlanguage}{english}
\begin{abstract}
\noindent
A new class of representations of the Brauer algebra that centralizes the action of orthogonal and symplectic groups in tensor spaces is found.
These representations make it possible to apply the technique of building primitive orthogonal
idempotents of the Brauer algebra to the construction of
integer spin Behrends-Fronsdal type projectors of an arbitrary type of symmetries.
\end{abstract}
\end{otherlanguage}

\vspace{3cm}

\newpage

\section{Introduction}
\setcounter{equation}{0}

It is thought that the
 theory of particles with higher spins  $j \, (j>2)$
began with the paper by Fierz and Pauli \cite{PF}.
Their approach was based on the imperative physical
requirements of the Lorentz invariance and positivity of energy (after quantization).
Further, from the works by Wigner \cite{Wig} and
Bargman and Wigner \cite{BWig} it became
 clear that these conditions can be replaced by the
requirement that single-particle states in quantum
field theory are described by special irreducible unitary representations
of the Poincar\'{e} group.
For example, a particle with integer spin $j$ and mass {\sf m} in
the momentum representation is described by the completely
symmetric rank $j$
tensor field $\phi (k) \in (\mathbb{R}^{1, 3})^{\otimes j}$
with the components satisfying the relations
 \be
 \lb{intr01}
 k_{n_ {\ell}} \phi^{n_1 \cdots n_{\ell} \cdots n_j} (k) = 0\; , \;\;\;
 \eta_{n_i n_{\ell}} \phi^{...n_i ...n_{\ell}...}(k) = 0  \; ,
 \ee
 where $k^n$ are the coordinates of the four-momentum $k \in \mathbb{R}^{1,3}$,
 the matrix $\eta = diag(1,-1,-1,-1)$
 is the metric of the four-dimensional Minkowski space, and
 the on-shell condition $k^2 \equiv k^n k_n = {\sf m}^2$ is implied.
Equations (\ref{intr01}) are called the {\bf \em transversality
and tracelessness conditions}, respectively.

To construct spin $j$ irreducible unitary representations of the
Poincar\'{e} group or in other words to extract fields obeying (\ref{intr01})
from tensor space $(\mathbb{R}^{1, 3})^{\otimes j}$,
one can use special operators.
These operators and their various generalizations
are usually called the spin projection operators or TT-projectors
(according to the first letters of the words
{\it tracelessness and transversality}, see (\ref{intr01})).

The first well-known examples of this type projectors
are the  Behrends-Fronsdal (BF) spin operators $\Theta^{(j)}$,
constructed in papers \cite{Fronsd}, \cite{BF},
devoted to the fundamentals of the theory of higher spin fields
(about this theory see, e.g., papers
\cite{SG}, \cite{GJ}, \cite{Vas}, \cite{Sor}, 
\cite{GolVas}, \cite{Gio} and references therein).
The operators $\Theta^{(j)}$ project the space of
rank $j$ tensors $\phi(k) \in  (\mathbb{R}^{1, 3})^{\otimes j}$
  onto invariant (under the action of the Poincar\'{e}
  group $ISO(1,3)$) subspace of the completely
 symmetric rank $j$ tensors $\phi(k)$ with the components satisfying
(\ref{intr01}). We note that the BF projectors proved extremely useful
 in elementary particle phenomenology. In particular, they are
 helpful for a systematic derivation
 of decay process amplitudes of particles with spin
within the framework of the tensor formalism
(see \cite{Zem}, \cite{Chu1} and \cite{Chu2}). 
 The BF projectors are also needed in the covariant tensor formalism
to obtain reliable results in the phenomenological
spin-parity description of resonances in the meson spectroscopy \cite{FFRo}.

A $D$-dimensional ($D>4$) generalization of the Behrends-Fronsdal projectors
for the case of integer spins was found for the first time in \cite{SEG}
(see also \cite{PoTs}, \cite{IP}).
The $D$-dimensional BF spin projection operator for half-integer spins
was constructed in \cite{IP} and \cite{IP1}. The $D$-dimensional
operators of this type are used in different fields of 
 modern theoretical physics\footnote{TT-projectors are used in
 the formulation of conformal
 higher-spin  actions proposed by Fradkin and Tseytlin \cite{FC}. For example, see
 the work \cite {BKFP}, where the TT-projectors are used to construct
higher-spin Cotton tensor controlling conformal geometry in three dimensional space-time (see also \cite{MHa}, \cite{K} and \cite{KM}).};
therefore, different forms of these projectors appear in 
the literature (see, e.g., \cite{PoTs}, \cite{IP2},  \cite{PM}).

 In this paper, we give a constructive way
 of building BF projectors onto the spaces of all infinite-dimensional
 irreducible tensor represen\-tations of the Poincar\'{e} group $ISO(1,D-1)$ for the case $D >4$. We stress that here
  we consider only tensor representations, which
  are analogs of integer spin representations for the $D=4$ case.
  The corresponding projectors act in the space
 $(\mathbb{R}^{1, D-1})^{\otimes j}$ and extract invariant subspaces
of rank $j$ tensors with the components $\Phi^{\ell_1...\ell_j}(k)$
(where $k \in \mathbb{R}^{1, D-1}$ is $D$-dimensional
 momentum) which are symmetrized under special permutations
 of indices $\ell_1,...,\ell_j$
 and satisfy $D$-dimensional generalization of the
 conditions (\ref{intr01}).

 So our aim is to construct explicitly projectors
 onto the spaces of TT-tensors of rank $j$
 with different types of symmetries that are spaces of
 irreducible representations of the group $ISO(1,D-1)$ for the
 multidimensional case $D >4$.
 It is known (see, e.g., \cite{IR} and references therein)
  that such symmetries can be
 associated with the Young diagrams with $j$ boxes.
 In particular, the completely symmetric
 rank $j$ tensors are associated with the one-row Young diagram with $j$
 boxes.
In the case of orthogonal groups $SO$, the
 finite dimensional irreducible tensor representa\-tions can be constructed
by means of the Schur--Weyl--Brauer duality. The key ingredients here are
the Brauer algebra and its primitive orthogonal idempotents (see
\cite {IR} and \cite {NYM}). The projectors onto invariant
 subspaces of rank $j$ tensors with special type
 of symmetries (spaces of irreducible
 tensor represen\-tations of the group $SO(1,D-1)$)
 are realized as images of the primitive orthogonal idempotents
 of the Brauer algebra ${\cal B}r_j$ acting in tensor spaces
 $(\mathbb{R}^{1, D-1})^{\otimes j}$. These images
  are analogs of the Young symmetrizers,
 which solve a similar problem for linear groups 
 $SL$ and $SU$.

As mentioned above, the irreducibility of
 tensor representations of the Poincar\'{e} group $ISO(1,D-1)$ requires
transversality and traceless properties of tensors that form 
 represen\-tation spaces.
In this paper, we show that the $D$-dimensional
Behrends-Fronsdal TT-projector of rank $j$ is nothing but a complete
 symmetrizer in the Brauer algebra ${\cal B}r_j$ taken in a special representation, which is described in Section 4.
 Making use of this fact, we
propose a fairly general algebraic method of
 constructing all other $D$-dimensional TT-projectors
 that possess an arbitrary type of symmetry.
Our method is based on the known constructions
of all Brauer algebra idempotents
 (see \cite{NYM}, \cite{IM},  \cite{IMO}) and on
 using of a new family of representations of the Brauer algebra
 (see Section 4).
 The images of idempotents of the Brauer algebra
 in these new represen\-tations automatically give
 TT-projectors.

The work is organized as follows. In the second section, we recall the
 definition of the Behrends-Fronsdal spin projection
 operators in the $D$-dimensional case.
 In the third section, we present a brief definition of
 the Brauer algebra. We introduce a system of its generators and
 give defining relations for these generators. Then we briefly
expose the procedure of constructing primitive orthogonal idem\-potents
of the Brauer algebra. In the fourth section,
we construct a new class of representations of the Brauer algebra
${\cal B}r_j(\omega)$ acting in the space of rank $j$ tensors.
Then, by using these representations, we present
 images of primitive idempotents corresponding to rank $j$
symmetrizers $\Theta_{{\{[j] ; j \}}}$
that are projected onto spaces of complete symmetric $j$-rank
tensors that are the spaces of irreducible represen\-tations
(of the group $ISO(1,D-1)$) associated with the $1$-row Young diagrams
with $j$ boxes. We prove that these symmetrizers
are equal to the $D$-dimensional
Behrends-Fronsdal spin $j$ projectors $\Theta^{(j)}(k)$
constructed in \cite{PoTs} and \cite{IP}.
In the fifth section, we describe another approach  to
construct the completely symmetric
projectors $\Theta_{{\{[j] ; j \}}}$. This
 approach is based on using the Zamolodchikov solution \cite{Zam}
 of the Yang-Baxter equations and
 gives the different construction \cite{IM}
  of the complete symmetrizers (\ref{rS1}) in the Brauer algebra $\mathcal{B}r_{j}$.
  In this section, we also obtain new recurrence relations
  for the $D$-dimensional
Behrends-Fronsdal spin $j$ projectors $\Theta^{(j)}(k)$.
In the sixth section,
by using the new Brauer algebra representations constructed
in Sect.4, we present a few examples of
images of primitive idempotents. These images are 
 TT-projectors onto
tensor spaces  of irreducible represen\-tations
of the group $ISO(1,D-1)$ associated with the special Young diagrams
$\lambda$: $[1^m]$ and $[2,1]$
(note that the number of rows $m$ in $\lambda$ can not exceed $r$, 
where $(D-1) = 2r,2r+1$).

 \section{Behrends-Fronsdal spin projection operator in the general case \label{BF-proj}}

 Here we recall
 the definition \cite{PoTs,IP}
 of the generalized $D$-dimensional Behrends-Fronsdal projector
 onto the spaces of irreducible completely symmetric tensor representations
 of the group $ISO(p,q)$, where $p+q=D$.

 \vspace{0.2cm}

 \noindent
 {\bf Definition 1.}
 {\it The operator $\Theta^{(j)}(k)$ in the space
 $(\mathbb{R}^{p, q})^{\otimes j}$, where $(p+q) = D$ and
 $k \in \mathbb{R}^{p, q}$, with the matrix
 $(\Theta^{(j)})^{n_1 \cdots n_j}_{r_1 \cdots r_j}(k)$
 is called the $D$-dimensional Behrends-Fronsdal projector
 if $\Theta^{(j)}$ has the following properties: \newline

\noindent
1) Projective property and hermiticity: $\;\;\;\;\;(\Theta^{(j)})^2=\Theta^{(j)},\;\; (\Theta^{(j)})^\dagger=\Theta^{(j)}$. \newline

\noindent
2) Complete symmetry:	\;\;\;
$(\Theta^{(j)})^{\; n_1\cdots\cdots n_j}_{\cdots r_i \cdots r_\ell \cdots}
=(\Theta^{(j)})^{\; n_1\cdots\cdots n_j}_{\cdots r_\ell \cdots r_i \cdots},\;\;
(\Theta^{(j)})^{\cdots n_i\cdots n_\ell \cdots}_{\; r_1 \cdots\cdots r_j}=
(\Theta^{(j)})^{\cdots n_\ell \cdots n_i\cdots}_{\; r_1 \cdots\cdots r_j}$. \newline

\noindent
3) Transversality:\;\;\;\;\;$k^{r_1}(\Theta^{(j)})^{n_1 \cdots n_j}_{r_1 \cdots r_j}=0$,
$\;\;k_{n_1}(\Theta^{(j)})^{n_1 \cdots n_j}_{r_1 \cdots r_j}=0$. \newline

\noindent
4) Tracelessness:\;\;\;
 $\eta^{r_1 r_2}(\Theta^{(j)})^{n_1 \cdots n_j}_{r_1 r_2\cdots r_j}=0$,
 $\eta_{n_1 n_2}(\Theta^{(j)})^{n_1 n_2 \cdots n_j}_{r_1 \cdots r_j}=0$}. \newline

\noindent
Here
 $$
 \begin{array}{c}
 ||\eta_{kl}|| = ||\eta^{kl}|| =
 {\rm diag}(\underbrace{+1,...,+1}_p,\underbrace{-1,...,-1}_q)  \; ,
 \end{array}
 $$
 is the metric in the space $\mathbb{R}^{p,q}$. We note that
 for real matrices $\Theta^{(j)}$ the hermiticity condition in 1.)
 is represented as $(\Theta^{(j)})^{n_1 \cdots n_j}_{r_1 \cdots r_j}=
 (\Theta^{(j)})_{n_1 \cdots n_j}^{r_1 \cdots r_j}$ and the second
 equations in 2.) -- 4.) follow from the first.

Instead of the matrix components
$(\Theta^{(j)})^{n_1 \dots n_j}_{r_1 \dots r_j}$
symmetrized in the upper and lower indices it is convenient to
 consider the generating function
 \be
 \lb{genT01}
 \Theta^{(j)}(x,u) =  u_{n_1} \cdots u_{n_j} \,
 (\Theta^{(j)})^{n_1 \dots n_j}_{r_1 \dots r_j} \,
 x^{r_1} \cdots x^{r_j} \; .
 \ee
where $x^{r}$ and $u^n$ are the components of the vectors
 $x,u \in \mathbb{R}^{p,q}$ and $u_r = \eta_{rn} u^n$.

\begin{proposition}\label{svop1}
(\cite{PoTs,IP}) The components of the $D$-dimensional spin projection operator
 $\Theta^{(j)}$ are defined uniquely by properties 1)-4) in Definition {\bf 1},
 and their generating function (\ref{genT01}) has the form
 \be
 \lb{genT02}
  \Theta^{(j)}(x,u) = \sum_{A=0}^{[\frac{j}{2}]} a^{(j)}_A \;
 \bigl(\Theta^{(u)}_{(u)} \, \Theta^{(x)}_{(x)} \bigr)^A  \;
 \bigl(\Theta^{(u)}_{(x)} \bigr)^{j -2A}  \; ,
 \ee
 where $[\frac{j}{2}]$ is the integer part of $j/2$,
  the coefficients $a^{(j)}_{A}$ (for $A=0$  and $A \geq 1$) are
  \be
 \lb{genT11}
 a^{(j)}_{0}=1 \, , \;\;\;\;
 a^{(j)}_{A} = \frac{(-1/2)^A \,j!}{
 (j -2A)! \, A! \, (2j +D-5)(2j +D-7)\cdots (2j +D -2A -3)}  \; ,
 \ee
 and $\Theta^{(u)}_{(x)}=\Theta^{(x)}_{(u)}$ denotes the function
 \be
 \lb{genT}
 \Theta^{(u)}_{(x)} \equiv  \Theta^{(1)}(x,u) =
 x^r \, u_n \, \Theta^{n}_{r}(k) \; ,  \;\;\;\;\;\;
 \Theta^{n}_{r}(k)  \equiv \delta^n_{r}-\frac{k_r k^n}{k^2} \; .
 \ee
\end{proposition}
{\bf Remark.} Constants (\ref{genT11}) satisfy the recurrence relations
 \be
 \lb{genT10lk}
 a^{(j)}_{A} = - \frac{1}{2} \; \frac{(j -2A+2)(j -2A+1)}{
 A \; (2j -2A+D-3)} \; a^{(j)}_{A-1} =
 - \frac{1}{2} \; \frac{(j -2A+1)j}{
 A \; (2j +D-5)} \; a^{(j-1)}_{A-1}\; ,
 \ee
 which are used below.
\begin{proposition} \lb{CH0}
For the generation function (\ref{genT01}) the following recurrence relation
 holds:
\be \lb{shreq4}
\begin{array}{c}
\Theta^{(j)}(x,u) = \frac{1}{(j-1)!}
\Bigl ( \Theta^{(x)}_{(u)}
- \,\, \frac{1}{(\omega + 2 (j-2))} \, \Theta^{(x)}_{(x)} \,
(u_k \, \partial_{x_k}) \Bigr)\,
\bigr ( \Theta^{(\,x\,)}_{(\partial_z)} \bigl )^{j-1}
\, \Theta^{(j-1)} (z,u)\,,
\end{array}
\ee
where  $\partial_{x_k} = \frac{\partial}{\partial x_k}$, the function $\Theta^{(x)}_{(u)}$ is defined in (\ref{genT}) and $\omega = (D-1)$.
\end{proposition}
{\bf Proof.} In the right-hand side of (\ref{shreq4}) the
differential operator
$\bigr ( \Theta^{(\,x\,)}_{(\partial_z)} \bigl )^{j-1}$ acts on
the generating function $\Theta^{(j-1)}(z,u)$,
which is given in (\ref{genT02})
where the coefficients $a_A^{(j-1)}$ are defined in (\ref{genT11}).
The result of this action is
\be \lb{Nsh0}
\begin{array}{c}
( \Theta^{(\,x\,)}_{(\partial_z)} \bigl )^{j-1} \, \Theta^{(j-1)} (z,u) = \displaystyle \sum_{A=0}^{[\frac{j-1}{2}]} a_A^{(j-1)}
\bigl(\Theta^{(u)}_{(u)}\bigr )^A \, \bigl ( \Theta^{(\,x\,)}_{(\partial_z)} \bigr )^{j-1} \, \Bigl ( \bigl ( \Theta^{(z)}_{(z)} \bigr)^A  \;
 \bigl(\Theta^{(u)}_{(z)} \bigr)^{j-1-2A} \Bigr ) \;  = \\ [0.5cm]
 = \, (j-1)! \, \displaystyle{ \sum_{A=0}^{[\frac{j-1}{2}]} a_A^{(j-1)}
\bigl(\Theta^{(u)}_{(u)} \, \Theta^{(x)}_{(x)} \bigr)^A  \;
 \bigl(\Theta^{(u)}_{(x)} \bigr)^{j-1-2A}} \, ,
\end{array}
\ee
where the second equality in (\ref{Nsh0}) follows from
 the formula
\be \lb{Nsh1}
\begin{array}{c}
\bigl ( \Theta^{(\,x\,)}_{(\partial_z)} \bigr )^{j-1} \, \Bigl ( \bigl ( \Theta^{(z)}_{(z)} \bigr)^A  \;
 \bigl(\Theta^{(u)}_{(z)} \bigr)^{j-1-2A} \Bigr ) \; =
 (j-1)! \bigl ( \Theta^{(x)}_{(x)} \bigr)^A \bigl(\Theta^{(u)}_{(x)} \bigr)^{j-1-2A}.
\end{array}
\ee
Now making use of (\ref{Nsh0}), we write the right-hand
side of (\ref{shreq4}) in the form
\be \lb{shs3}
\begin{array}{c}
 a_0^{(j-1)} \, \bigl(\Theta^{(u)}_{(x)} \bigr)^{j} +
 \sum\limits_{A=1}^{[\frac{j-1}{2}]} B_A^{(j)}
\bigl(\Theta^{(u)}_{(u)} \, \Theta^{(x)}_{(x)} \bigr)^A \;
 \bigl(\Theta^{(u)}_{(x)} \bigr)^{j-2A} \, - \\ [0.4cm]
- \left. \frac{(j+1-2A)}{(\omega + 2(j-2))} \cdot a_{A-1}^{(j-1)}
\bigl(\Theta^{(u)}_{(u)} \, \Theta^{(x)}_{(x)} \bigr)^{A}  \;
 \bigl(\Theta^{(u)}_{(x)} \bigr)^{j-2A} \right|_{A=[\frac{j-1}{2}]+1} \; ,
\end{array}
\ee
where the coefficients $B_A^{(j)}$ are determined as
follows:
\be \lb {shs5}
\begin{array}{c}
B_A^{(j)} = a_A^{(j-1)} - \frac{2A}{(\omega + 2(j-2))} a_A^{(j-1)} - \frac{(j-2A+1)}{(\omega + 2(j-2))} a_{A-1}^{(j-1)} = a_{A}^{(j)} \; .
\end{array}
\ee
Here, in the last equality, we have used the
recurrence relations (\ref{genT10lk})
for the coefficients $a^{(j)}_{A}$ and condition $\omega = (D-1)$. Note that
for odd and even $j$ we have respectively the relations
$[\frac{j-1}{2}] = [\frac{j}{2}]$ and $[\frac{j-1}{2}] = [\frac{j}{2}]-1$.
Besides,  for odd $j$, the last term in (\ref{shs3}) is zero, while
 for even  $j$ this term is equal to
 $$
 \left. a_A^{(j)}
\bigl(\Theta^{(u)}_{(u)} \, \Theta^{(x)}_{(x)} \bigr)^A  \;
 \bigl(\Theta^{(u)}_{(x)} \bigr)^{j-2A} \right|_{A =  [\frac{j}{2}]} \; .
 $$
As a result,
the whole expression in (\ref{shs3}) can be written
(for both even and odd $j$) as
\be \lb{shs6}
 \displaystyle{\sum_{A=0}^{[\frac{j}{2}]} a_A^{(j)}
\bigl(\Theta^{(u)}_{(u)} \, \Theta^{(x)}_{(x)} \bigr)^A  \;
 \bigl(\Theta^{(u)}_{(x)} \bigr)^{j-2A}} \equiv
 \Theta^{(j)}(x,u) \; ,
\ee
which proves the identity (\ref{shreq4}). \hfill \qed

\section{Brauer algebra and its idempotents} \lb{baa}

In this section, we construct primitive idempotents in the Brauer algebra.
 We use these idempotents in the next sections for building 
 TT-projectors, which act in tensor spaces of irreducible 
 representations of the group $ISO(1,D-1)$.
Here we follow the exposition of \cite{IR}, which is based on the
results of papers \cite{NYM}, \cite{OV}
 (see also \cite{IM,IMO} and references therein).

 \vspace{0.2cm}

 \noindent
 {\bf Definition 2.}
{\it The unital associative algebra $\mathcal{B}r_j(\omega)$ over the field of
 complex numbers with generators $\sigma_i$ and $\kappa_i$ ($i=1, \dots, j-1$)
and defining relations (see \cite{BR} and, e.g., \cite{NYM,IM})
\be \label{drel}
\begin{array}{c}
\sigma_i^2 = e, \;\;\; \kappa_i^2 = \omega \kappa_i, \;\;\; \sigma_i \kappa_i = \kappa_i \sigma_i = \kappa_i, \;\;\; i = 1, \dots, j-1,  \\[0.5cm]
\sigma_i \sigma_{\ell} = \sigma_{\ell} \sigma_i, \;\;\; \kappa_i \kappa_{\ell} = \kappa_{\ell} \kappa_i, \;\;\; \sigma_i \kappa_{\ell} = \kappa_\ell \sigma_i, \;\;\; |i-\ell| > 1, \\[0.5cm]
\sigma_i \sigma_{i+1} \sigma_i = \sigma_{i+1} \sigma_i \sigma_{i+1},  \;\;\; \kappa_i \kappa_{i+1} \kappa_i =  \kappa_i  \;\;\; \kappa_{i+1} \kappa_i \kappa_{i+1} = \kappa_{i+1},  \\[0.5cm]
\sigma_i \kappa_{i+1} \kappa_i = \sigma_{i+1} \kappa_i,  \;\;\; \kappa_{i+1} \kappa_i  \sigma_{i+1}= \kappa_{i+1} \sigma_i ,  \;\;\; i = 1, \dots, j-2,
\end{array}
\ee
is called the {\bf \em Brauer algebra}. Here $e$ is a unit element
and $\omega$ is a real parameter characterizing the algebra.}

\vspace{0.2cm}

All basis elements of the algebra $\mathcal{B}r_j(\omega)$
are composed as products of the generators
$\sigma_i$ and $\kappa_i$. The dimension of the
 Brauer algebra $\mathcal{B}r_j(\omega)$ is
\be \lb{diB}
dim(\mathcal{B}r_j) = (2n-1)!! = (2n-1)(2n-3)\cdots 3 \cdot 1 \, .
\ee
One can consider the algebra  $\mathcal{B}r_j(\omega)$
as an extension of the group algebra $\mathbb{C}[S_j]$
of the permutation group $S_j$. The algebra  $\mathcal{B}r_j(\omega)$ plays the same role in the
theory of representations of the orthogonal
groups $SO(N, \mathbb{C})$ (and their real forms $SO(p,q)$)
as the group algebra $\mathbb{C}[S_j]$ in the theory
of representations of linear groups $SL(N, \mathbb{C})$
(and their real forms $SU(N)$).

Define \cite{IR,NYM} a set of special elements $y_m \in \mathcal{B}r_j(\omega)$
 $(m = 1, \dots j )$
\begin{equation}\lb{jme1}
\begin{array}{c}
\displaystyle
y_1 = 0, \;\;\; y_m  = \sum_{k=1}^{m-1}  (\sigma_{k,m} - \kappa_{k, m}) = \\[0.5cm]
\displaystyle
= \sum_{k=1}^{m-1}  \sigma_{m-1} \cdots \sigma_{k+1} (\sigma_k - \kappa_k) \sigma_{k+1} \cdots \sigma_{m-1} \, , \;\;\; m = 2, 3, \dots , j \, ,
\end{array}
\end{equation}
where
\be \lb{Sh}
\begin{array}{c}
\sigma_{k,m} = \sigma_{m-1} \cdots \sigma_{k+1} \sigma_k \sigma_{k+1} \cdots \sigma_{m-1},
 \;\;\; \kappa_{k, m} =  \sigma_{m-1} \cdots \sigma_{k+1} \kappa_k \sigma_{k+1} \cdots \sigma_{m-1} \, .
\end{array}
\ee
The operators $y_m \in \mathcal{B}r_j(\omega)$ are called the Jucys-Murphy elements
and play an important role in
 constructing primitive orthogonal idempotents of $\mathcal{B}r_j(\omega)$.
Note that the Jucys-Murphy elements can be
 expressed via the recurrence relation
\begin{equation}\lb{jme2}
y_1 = 0, \;\;\;  y_{n+1} = \sigma_n - \kappa_n + \sigma_n y_n \sigma_n \, ,
\end{equation}
 and for illustration we present the first two nontrivial elements
 \begin{equation}\lb{jme25}
y_2 = \sigma_1 - \kappa_1, \;\;\;  y_{3} = \sigma_2 - \kappa_2 +
\sigma_2 (\sigma_1 - \kappa_1)  \sigma_2 \, .
\end{equation}
We define a subalgebra $Y_j \in \mathcal{B}r_j(\omega)$
generated by all Jucys-Murphy elements $\{y_1, y_2, \dots y_j \}$.
For brevity, below we omit $\omega$ in the notation
$\mathcal{B}r_j(\omega)$ and write $\mathcal{B}r_j$.
The following statement holds (see \cite{IR},\cite{NYM}).
\begin{proposition} \lb{mcs0}
The Jucys-Murphy elements, defined in (\ref{jme1}), form a
complete set of commuting generators in $\mathcal {B}_j$																														
\be \lb{mcs}
[y_i, y_{\ell}] = 0, \;\;\; \forall i, \ell \, .
\ee
The algebra $Y_j$ is a maximal commutative subalgebra in $\mathcal {B}_j$.
\end{proposition}

Now we briefly discuss the procedure for constructing
a complete system of primitive orthogonal idempotents
$e_{\alpha} \in \mathcal{B}r_j$ that satisfy the relations
\be \lb{defPE}
e_{\alpha} e_{\beta} = \delta_{\alpha \beta} e_{\alpha},
\;\;\;\;\;\;\;  \displaystyle {\sum_{\alpha} e_{\alpha} = 1}.
\ee
 In addition,
 we require, in the left regular representation of
 $\mathcal{B}r_j$, the elements $e_{\alpha} \in \mathcal{B}r_j$
 to be eigenvectors of the Jucys-Murphy generators:
\be \lb{ev1}
y_m e_{\alpha} = a_{m}^{_{(\alpha)}} e_{\alpha}, \;\;\;\;\;\;\;\;
  a_{m}^{_{(\alpha)}} \in \mathbb{R}, \;\;\; \forall m = 1, 2, \dots, j \; .
\ee
Such a choice of idempotents
 $e_{\alpha} \in \mathcal{B}r_j$ is always possible, since
Jucys-Murphy elements $\{ y_1, \dots , y_j \}$ commute with each other. Moreover,
it can be shown (see \cite{IR}) that the idempotents
 $e_{\alpha}$ satisfying (\ref{defPE}) and (\ref{ev1}) commute with all Jucys-Murphy elements
 $y_m$ (it means that $e_{\alpha}$ are the functions of $y_m$) and
all eigenvalues $a_{m}^{_{(\alpha)}}$ in (\ref{ev1}) are real numbers.
 According to (\ref{ev1}),
each primitive idempotent $e_{\alpha} \in \mathcal{B}r_j$ is
 characterized by a set of
 eigenvalues $a_{i}^{_{(\alpha)}}$ that form the spectral vector
\be \lb{spSet}
\Lambda_{\alpha} = ( a_1^{_{(\alpha)}}, a_2^{_{(\alpha)}}, \dots , a_{j}^{_{(\alpha)}}) \in \mathbb{R}^{j} \, ,
\ee
 where in view of (\ref{jme1}) we have $a_1^{_{(\alpha)}} = 0$.
 We denote the set of all spectral vectors (\ref{spSet}) as
 $Spec (y_1, \dots , y_n)$. It is possible to prove
 (the proof is given in $\cite{IR}$)
 that eigenvalues $a_{i}^{_{(\alpha)}}$ of $y_i$ satisfy the condition
\be \lb{sJM}
a_{i}^{_{(\alpha)}} \in \{ [1-i, i-1], [2-i, i-2] + (1 - \omega) \} ,
\ee
where the bracket $[-z,z]$ denotes a set of integers
\be \lb{sJM1}
[-z, z] = \{-z, \dots, -1,0,1, \dots, z\}, \;\;\; z \in \mathbb{Z}_{\geq 0} \, ,
\ee
and $[-z,z] +a$ denotes a set of integers shifted by $a$
\be \lb{sJM2}
[-z, z] + a= \{a-z, \dots, a-1,a,a+1, \dots, a+z\} \, .
\ee
The remaining conditions that completely determine the elements of the set
 $Spec (y_1, \dots y_j)$ can be found in \cite{IR,NYM}.

\begin{proposition} \lb{corr} Elements of the set $Spec(y_1, \dots y_j)$
correspond one-to-one to the elements
of a set of oscillating Young tableaux.
\end{proposition}

In the formulation of this Proposition, we use the notion of the
{\bf \em oscillating Young tableau}. Now we recall (see, e.g., \cite{IR,IM,IMO})
the notions of the Young diagrams and oscillating Young tableaux.
The Young diagram $\lambda$ with $r$ boxes and $k$ rows
 is a set of integers $\{ m_1, m_2,...,m_k \}$ such that $m_1 \geq m_2 \geq ...
\geq m_k >0$ and $\sum_{i=1}^k m_k = r$. The standard notation is
$\lambda = [m_1, m_2,...,m_k] \vdash r$.
 Consider a sequence of Young diagrams ${\sf \Lambda} =
 \{\lambda_0, \lambda_1, \dots , \lambda_j\}$,
which starts with a trivial  diagram $\lambda_0 = \emptyset$, and the
diagram $\lambda_{k+1}$
standing in the sequence $\Lambda$ after $\lambda_k$ is obtained
either by adding one box to
the outer angle of the diagram $\lambda_k$ or by deleting
one box in the inner angle of the diagram
$\lambda_k$.  Such a sequence ${\sf \Lambda}$ of Young diagrams
is called the {\bf \em oscillating Young tableau}
(or updown Young tableau) of length $j$.
In the figure below, we give an example of possible transitions from
$\lambda_k$ to $\lambda_{k+1}$ in the oscillating Young tableau.
In the first line we depict the Young diagram $\lambda_{k}^{_{(a)}}=[3,2,2]$.
The second line contains five diagrams that can be obtained from
$\lambda_{k}^{_{(a)}}$ by the operations,
which were described above (adding or removing one box).

 \begin{figure}[h!]
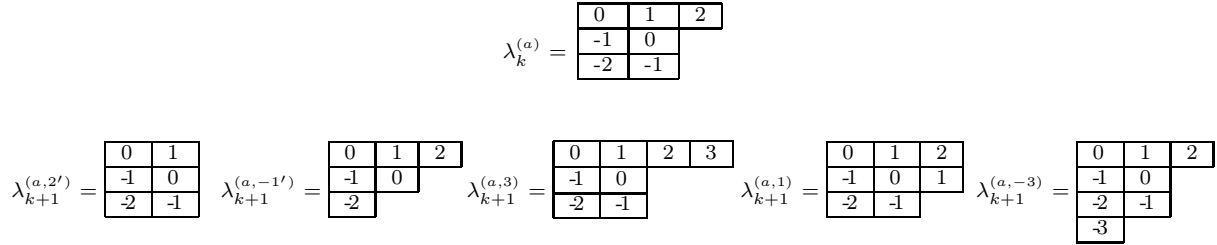

{\scriptsize
\begin{center}
$\lambda_{k}^{_{(a)}} = \,\,\,$\begin{tabular}{llll}
\cline{1-3}
\multicolumn{1}{|l|}{0}  & \multicolumn{1}{l|}{1}  & \multicolumn{1}{l|}{2} &  \\ \cline{1-3}
\multicolumn{1}{|l|}{-1} & \multicolumn{1}{l|}{0}  &                        &  \\ \cline{1-2}
\multicolumn{1}{|l|}{-2} & \multicolumn{1}{l|}{-1} &                        &  \\ \cline{1-2}
                         &                         &                        &
\end{tabular}
\end{center}

\begin{center}
$\lambda_{k+1}^{_{(a,2')}} = \,\,$\begin{tabular}{llll}
\cline{1-2}
\multicolumn{1}{|l|}{0}  & \multicolumn{1}{l|}{1}  &  &  \\ \cline{1-2}
\multicolumn{1}{|l|}{-\!1} & \multicolumn{1}{l|}{0}  &  &  \\ \cline{1-2}
\multicolumn{1}{|l|}{-\!2} & \multicolumn{1}{l|}{-\!1} &  &  \\ \cline{1-2}
                         &                         &  &
\end{tabular}
\!\!\!\!\!\!\!\!\!\!\!\!\!\!\!
$\lambda_{k+1}^{_{(a,-1')}} = \,\,$\begin{tabular}{llll}
\cline{1-3}
\multicolumn{1}{|l|}{0}  & \multicolumn{1}{l|}{1} & \multicolumn{1}{l|}{2} &  \\ \cline{1-3}
\multicolumn{1}{|l|}{-\!1} & \multicolumn{1}{l|}{0} &                        &  \\ \cline{1-2}
\multicolumn{1}{|l|}{-\!2} &                        &                        &  \\ \cline{1-1}
                         &                        &                        &
\end{tabular}
\!\!\!\!\!\!\!\!\!\!\!
$\lambda_{k+1}^{_{(a,3)}} = \,\,$\begin{tabular}{llll}
\hline
\multicolumn{1}{|l|}{0}  & \multicolumn{1}{l|}{1}  & \multicolumn{1}{l|}{2} & \multicolumn{1}{l|}{3} \\ \hline
\multicolumn{1}{|l|}{-\!1} & \multicolumn{1}{l|}{0}  &                        &                        \\ \cline{1-2}
\multicolumn{1}{|l|}{-\!2} & \multicolumn{1}{l|}{-\!1} &                        &                        \\ \cline{1-2}
                         &                         &                        &
\end{tabular}
\!\!
$\lambda_{k+1}^{_{(a,1)}} = \,\,$\begin{tabular}{llll}
\cline{1-3}
\multicolumn{1}{|l|}{0}  & \multicolumn{1}{l|}{1}  & \multicolumn{1}{l|}{2} &  \\ \cline{1-3}
\multicolumn{1}{|l|}{-\!1} & \multicolumn{1}{l|}{0}  & \multicolumn{1}{l|}{1} &  \\ \cline{1-3}
\multicolumn{1}{|l|}{-\!2} & \multicolumn{1}{l|}{-\!1} &                        &  \\ \cline{1-2}
                         &                         &                        &
\end{tabular}
\!\!\!\!\!\!\!\!\!
$\lambda_{k+1}^{_{(a,-3)}} = \,\,$\begin{tabular}{|l|lll}
\cline{1-3}
0  & \multicolumn{1}{l|}{1}  & \multicolumn{1}{l|}{2} &  \\ \cline{1-3}
-\!1 & \multicolumn{1}{l|}{0}  &                        &  \\ \cline{1-2}
-\!2 & \multicolumn{1}{l|}{-\!1} &                        &  \\ \cline{1-2}
-\!3 &                         &                        &  \\ \cline{1-1}
\end{tabular}
\end{center}

\begin{otherlanguage}{english}
\caption{\label{Fig0} {\it Examples of possible transitions between the diagrams $\lambda_k$ and $\lambda_{k+1}$ in an oscillating Young tableau. }}
\end{otherlanguage}
}
\end{figure}

\noindent
Here, in the notation of the Young diagrams $\lambda_k^{(a)}$, in
addition to the lower index $k$, we introduce  the upper multi-index
 $(a)=(a_1, a_2, \dots, a_k) \in \mathbb{R}^k$ which is a content vector
 of the oscillating Young tableau, where the last Young diagram is $\lambda_k$.
 Now using the tables from Fig.\ref{Fig0} as an example, we will explain according to what rules we select the value of the element $a_{k+1}$
 when moving from the table $\lambda_{k}^{_{(a)}}$ to the table $\lambda_{k+1}^{_{(a,a_{_{k+1}})}}$, that is, when passing from the vector
$(a) \in \mathbb{R}^k$ to the vector $(a,a_{k+1}) = (a_1, a_2, \cdots a_k, a_{k+1}) \in \mathbb{R}^{k+1}$.

Define the coordinates $(s,t) \in \mathbb{Z}_{+}^2$
of boxes of an arbitrary Young diagram $\lambda$,
where $s$ -- the row number and $t$ -- the column number of the diagram $\lambda$.
For example, the upper left angle of the diagram has the coordinates $s=1$ and $t=1$.
The value $(t-s)$ will be called the content of the box with the coordinates $(s,t)$.
The diagram boxes in Fig. {\bf \ref{Fig0}} are filled with the content, according to this rule.
In the general case, when the box with the coordinates $(s,t)$ is added to the diagram
$\lambda_{k}^{_{(a)}}$, then the coordinate $a_{k+1} = (t-s)$ is added to the vector $(a) \in \mathbb{R}^k$
of the new diagram $\lambda_{k+1}^{_{(a,a_{k+1})}}$ thus obtained, and if the box with the coordinates
$(s,t)$ is deleted, then the coordinate $a_{k+1} = (t-s)'= (1- \omega) - (t-s)$
is added to the vector $(a)$ in the notation of the new diagram.
For example, consider the diagram $\lambda_{k+1}^{_{(a,2')}}$
from the second line of Figure {\bf \ref{Fig0}}. It is obtained from
 $\lambda_{k}^{_{(a)}}$ by deleting the box with the content $2$;
 therefore, the value $(a_1, \dots, a_k)$
 is added to the vector $a_{k+1} = 2' = (1-\omega)-2$.

Thus, each oscillating Young tableau corresponds to a sequence of diagrams with transitions between them:
\be \lb{vco1}
\Lambda = \{ \emptyset \overset{a_1 = 0}{\rightarrow} \lambda_1  \overset{a_2}{\rightarrow}
\lambda_2  \overset{a_3}{\rightarrow} \lambda_3  \overset{a_4}{\rightarrow} \cdots
 \overset{a_j}{\rightarrow} \lambda_j \} \, ,
\ee
where each transition $ \lambda_{k} \rightarrow \lambda_{k + 1} $ is assigned a value
$ a_{k + 1} $ determined from the content value of the added or deleted box.
It follows that to each oscillating Young tableau  $ \Lambda$
of length $j$ there corresponds a vector called the {\it content vector} of length $j$
\be \lb{vco2}
A = (a_1, a_2, \dots a_j) \, ,
\ee
where, according to the rules for determining the numbers $a_i$ for added or removed boxes, we have
(see (\ref{sJM}))
\be \lb{sJM3}
a_i \in \{ [1-i, i-1], \, [2-i, i-2] + 1 - \omega \} \, .
\ee
The sequence set $(a_1, a_2, \dots, a_j)$ for all oscillating Young tableaux of length
 $j$  is called the set of content vectors of length $j$.
The elements of this set uniquely correspond to the oscillating Young tables, as indicated in Proposition {\bf \ref{corr}}.
On the other hand (see  \cite{IR}), the set of content vectors coincides with the set $Spec(y_1, \dots, y_j)$.

\begin{figure}[h!]
\setlength{\unitlength}{0.00023495in}%
\begingroup\makeatletter\ifx\SetFigFont\undefined%

\fi\endgroup%
\begin{picture}(15266,13019)(8668,-12168)

 \put(10000,-300){$y_1=$}
  \put(10000,-2100){$y_2=$}
   \put(10000,-4800){$y_3=$}
    \put(10000,-8400){$y_4=$}


\put(23550,-250){\scriptsize $0$}
\put(20100,-2000){\scriptsize $0'$}
\put(22900,-2000){\scriptsize $+1$}
\put(24500,-2000){\scriptsize $-1$}

\put(16000,-4800){\scriptsize $0$}
\put(18600,-4700){\scriptsize $-1'$}
\put(20300,-4800){\scriptsize $+1'$}

\put(22100,-5000){\scriptsize $+2$}

\put(23200,-5100){\scriptsize $-1$}
\put(24100,-4700){\scriptsize $+1$}
\put(25800,-4800){\scriptsize $-2$}


\put(14000,-8400){\scriptsize $0'$}
\put(15800,-8700){\scriptsize $+1$}
\put(16900,-8400){\scriptsize $-1$}
\put(18300,-8400){\scriptsize $-2'$}
\put(19900,-8400){\scriptsize $+1'$}

\put(20300,-9300){\scriptsize $-1'$}
\put(21600,-8700){\scriptsize $+3$}
\put(22500,-8300){\scriptsize $-1$}

\put(23200,-9900){\scriptsize $+2$}
\put(24400,-9500){\scriptsize $0$}

\put(25000,-8900){\scriptsize $-2$}
\put(25100,-7800){\scriptsize $+2'$}

\put(26400,-9100){\scriptsize $+1$}
\put(27300,-8400){\scriptsize $-3$}


 \put(23801,339){$\emptyset$}

\put(23476,-3061){\circle*{300}}
\put(23101,-3061){\circle*{300}}

 \put(22501,-6361){\circle*{300}}
 \put(22126,-6361){\circle*{300}} \put(22876,-6361){\circle*{300}}
\put(24001,-1111){\circle*{300}}

 \put(22501,-11836){\circle*{300}}
 \put(22501,-11461){\circle*{300}}
  \put(22876,-11461){\circle*{300}}
\put(23251,-11461){\circle*{300}}

 \thicklines

 \put(22501,-6661){\vector(0,-1){4500}}
  \put(24001,239){\vector( 0,-1){1050}}

 \put(23900,-1300){\vector( -1, -3){500}}

 \put(23176,-3436){\vector(-1,-4){675}}
  \put(25000,-3586){\vector(-1,-4){650}}
 \put(24451,-6961){\vector(1,-3){1500}}
 \put(26326,-6961){\vector( 0,-1){4425}}
 \put(24301,-6961){\vector( 0,-1){4425}}
 \put(24151,-6961){\vector(-1,-3){1350}}

 \put(22276,-6586){\vector(-1,-3){1425}}
 \put(16400,-6586){\vector( 1,-4){1050}}

 \put(16351,-6586){\vector(-1,-4){1050}}
 \put(25051,-3511){\vector( 1,-2){1200}}

 \put(23326,-3286){\vector(1,-3){975}}
 \put(16501,-3361){\vector( 0,-1){2700}}
 \put(24151,-1261){\vector( 1,-3){550}}

 \put(26626,-6811){\vector(1,-3){1200}}
 \put(24676,-3586){\vector(-3,-1){7852.500}}
  \put(24076,-6600){\vector(-3,-2){6529}}

 \put(23700,-1200){\vector( -4,-1){6900}}

 \put(22700,-3100){\vector( -2, -1){6100}}

 \put(26200,-7000){\vector( -2,-1){8500}}
 \put(16200,-6650){\vector( -1, -1){4000}}
 \put(24000,-6650){\vector( -2, -1){8400}}
 \put(21850,-6500){\vector( -3,-2){6525}}

 \put(12000,-11301){$\emptyset$}
 \put(16301,-3161){$\emptyset$}

  \put(21226,-11236){\circle*{300}}
 \put(24301,-11686){\circle*{300}}
\put(24301,-12061){\circle*{300}} \put(24676,-11686){\circle*{300}}
 \put(24676,-12061){\circle*{300}}
 \put(26101,-11761){\circle*{300}}
\put(26101,-12136){\circle*{300}} \put(26101,-12511){\circle*{300}}
\put(26476,-11761){\circle*{300}} \put(27901,-10636){\circle*{300}}
\put(27901,-11011){\circle*{300}} \put(27901,-11386){\circle*{300}}
\put(27901,-11761){\circle*{300}}  \put(20851,-11236){\circle*{300}}

   \put(24901,-2911){\circle*{300}} \put(24901,-3286){\circle*{300}}

\put(24301,-6361){\circle*{300}} \put(24676,-6361){\circle*{300}}
\put(24301,-6736){\circle*{300}} \put(26401,-6361){\circle*{300}}
\put(26401,-5986){\circle*{300}} \put(26401,-6736){\circle*{300}}
\put(16501,-6361){\circle*{300}}
\put(15500,-11161){\circle*{300}}\put(15100,-11161){\circle*{300}}
\put(17401,-11011){\circle*{300}}
\put(17401,-11386){\circle*{300}} \put(20101,-11236){\circle*{300}}
\put(20476,-11236){\circle*{300}}

\end{picture}

\begin{otherlanguage}{english}
\caption{\label{Fig1} {\it The colored oscillating Young graph for the Brauer algebra
 $\mathcal{B}r_4$. The indices assigned to the edges (arrows) of the graph are the eigenvalues of the Jucys-Murphy operators $y_k \in \mathcal{B}r_4$,
which are indicated on the left. }}
\end{otherlanguage}
\end{figure}
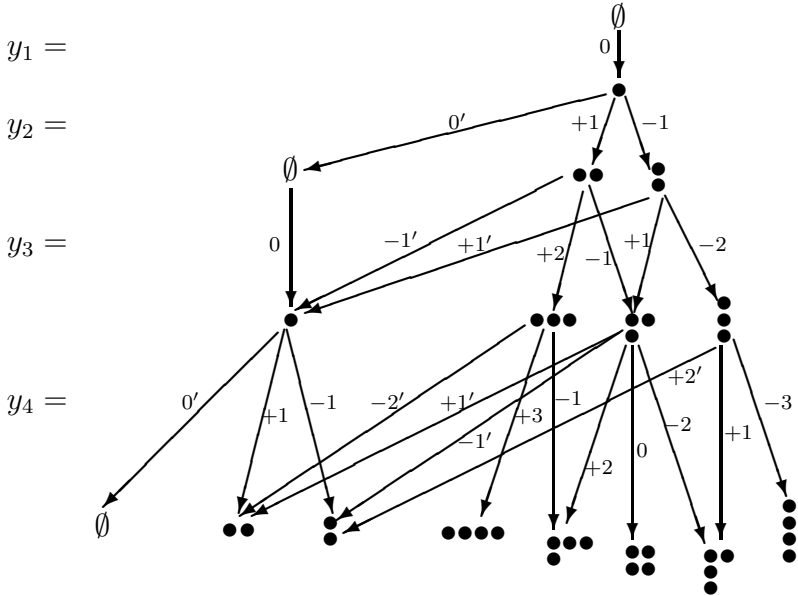

We introduce the concept of the colored oscillating 
 Young graph\footnote{''Colors'' in this case are eigenvalues of the Jucys-Murphy elements
$y_k$ assigned to the edges of the graph.
The idea to assign these eigenvalues to the edges
of the Young graph belongs to O.V.~Ogievetsky and one of the author
of this paper \cite{IO2}.}
 of the algebra $\mathcal{B}r_j$,
which is a convenient visual representation of all possible oscillating Young tableaux (\ref{vco1}) of fixed length $j$.
On the one hand, the vertices of such a graph at the level $j$ correspond to irreducible representations of the algebra $\mathcal{B}r_j$,
on the other hand, such a graph indicates the branching rules of these representations (that is, indicates possible transitions from the diagram at the level of $k$
to the diagrams at the level of  $(k+1)$).
Note that the dimension of the representation $\mathcal{B}r_k$, corresponding to the vertex $\lambda$
 at the level  $k$ of the Young graph is equal to the number of paths starting at the very top vertex $\emptyset$ and ending at this vertex $\lambda$.
  Thus, the oscillating Young graph encodes information on irreducible representations of the Brauer algebras, including the branching rules of these representations. For instance,
   Fig. {\bf \ref{Fig1}} (taken from the book \cite{IR})
   shows the oscillating Young graph for ${\cal B}_4$, which gives information on irreducible representations of the Brauer algebras ${\cal B}_k$, with $k=1,2,3,4$.

We now proceed to the derivation (see, for example, \cite{IR}) of general formulas for primitive orthogonal idempotents of the Brauer algebra.
The following figure shows a schematic view of the Young diagram  $\lambda$
located at the $n$-th level of the oscillating Young graph.

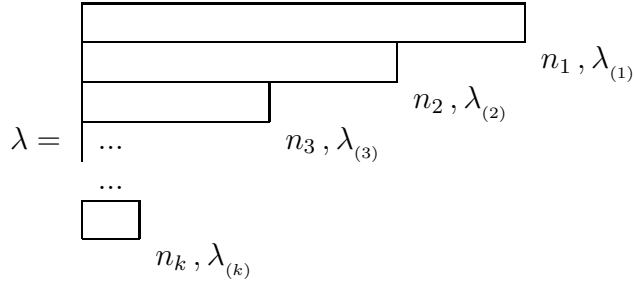
\begin{center}
$\lambda = \,\,$
\begin{tabular}{llllll}
\cline{1-4}
\multicolumn{4}{|l|}{}                           &        &  \\ \cline{1-4}
\multicolumn{3}{|l|}{}                   &       & $n_1\, , \lambda_{_{(1)}}$&  \\ \cline{1-3}
\multicolumn{2}{|l|}{}           &       & $n_2\, , \lambda_{_{(2)}} $&        &  \\ \cline{1-2}
\multicolumn{1}{|l}{...} &       & $n_3\, , \lambda_{_{(3)}}$&       &        &  \\
...                      &       &       &       &        &  \\ \cline{1-1}
\multicolumn{1}{|l|}{}   &       &       &       &        &  \\ \cline{1-1}
                         & $n_k\, , \lambda_{_{(k)}} $ &       &       &        &
\end{tabular}
\end{center}
\noindent
Here $(n_i, \lambda_{_{(i)}})$ are the coordinates of the boxes located in the inner angles of $\lambda$. Note that the number of boxes $|\lambda|$ of all diagrams $\lambda$
at the level $n$ obeys the inequality $|\lambda| \leq n$.
Consider in the oscillating Young graph of the algebra
 $\mathcal{B}r_{n+1}(\omega)$ any path $T_{\{\lambda;n\}}$
going down from the vertex $\{\emptyset; 0\}$ to the vertex
 of $\{\lambda; n\}$ \footnote{Here, in the notation for the vertex
 $\{\lambda; n\}$ the second character $n$ in braces indicates the level at which the diagram $\lambda$ appears. The level is indicated
 since in the oscillating Young graph (see Figure {\bf \ref{Fig1}}) the same diagrams can appear at different levels.}
, i.e., consider the path corresponding to the oscillating tableau $\Lambda = \{\lambda_1 = \emptyset, \lambda_2, \dots , \lambda_n = \lambda \}$.
Let $E_{T_{\{\lambda;n\}}} \in \mathcal{B}r_{n}(\omega)$ be
 a primitive orthogonal idempotent corresponding to
$T_{\{\lambda;n\}}$. Using the {\it branching rule},
  which is dictated by the oscillating Young graph
 for the Brauer algebra $\mathcal{B}r_{n+1}(\omega)$
 (see the example in Figure  {\bf \ref{Fig1}}),
 we can conclude that in order to move from the diagram
 $\lambda$ at the $n$ level to the diagram
$\lambda_{n+1}$ at the $(n+1)$ level along the path $T_{\{\lambda_{n+1};n+1\}}$, we need to add one box to the
 $\lambda$ diagram (with the content $(\lambda_{_{(r)}} - n_{r-1})$ ) to the outer angle of $\lambda$,
or remove one box (with the content of $(\lambda_{_{(r)}} - n_r)$), from the inner angle of the diagram $\lambda$.
Knowing the contents of the added or deleted boxes, we know all possible eigenvalues
$(\lambda_{_{(r)}} - n_{r-1}) $  or $(1-\omega + n_r - \lambda_{_{(r)}} )$ of the element $y_{n+1}$  in the representation defined by $E_{T_{\{\lambda;n\}}}$
and therefore we have the identity
\be \lb{harT}
E_{T_{\{\lambda;n\}}} \prod_{r=1}^{k+1}(y_{n+1} - (\lambda_{_{(r)}} - n_{r-1})) \prod_{r=1}^{k}(y_{n+1} - (1-\omega + n_r-\lambda_{_{(r)}})) = 0 \, ,
\ee
where we put  $\lambda_{_{(k+1)}} = n_0 =0.$ Thus, for the new diagram  $\lambda_{n+1} = \lambda'$
obtained by adding the box to the diagram $\lambda$ with the coordinates $(n_{i-1} +1, \lambda_{_{(i)}}+1)$, the corresponding primitive idempotent
(after suitable normalization) has the form \cite{IR}
\be \lb{EA}
\begin{array}{c}
E_{T_{\{\lambda';n+1\}}} =
E_{T_{\{\lambda;n\}}} \cdot
{\displaystyle \prod_{\overset{r=1}{r \neq i}}^{k+1}} \frac{(y_{n+1} - (\lambda_{_{(r)}} - n_{r-1}))}
{\bigl( (\lambda_{_{(i)}} - n_{i-1}) - (\lambda_{_{(r)}} - n_{r-1}) \bigr )}
 {\displaystyle \prod_{r=1}^{k}}\frac{(y_{n+1} - (1-\omega + n_r-\lambda_{_{(r)}}))}
{\bigl( (\lambda_{_{(i)}} - n_{i-1}) - (1-\omega  + n_{r} - \lambda_{_{(r)}}) \bigr )} \, .
\end{array}
\ee
For the new diagram $\lambda_{n+1} = \lambda''$ obtained from $\lambda$ by deleting the box with the coordinates
$(n_i, \lambda_{_{(i)}})$, we get a primitive idempotent \cite{IR}
\be \lb{ED}
\begin{array}{c}
E_{T_{\{\lambda'';n+1\}}} =
E_{T_{\{\lambda;n\}}} \cdot
{\displaystyle \prod_{r=1}^{k+1}} \frac{(y_{n+1} - (\lambda_{_{(r)}} - n_{r-1}))}
{\bigl( (1-\omega+n_i - \lambda_{_{(i)}}) - (\lambda_{_{(r)}}-n_{r-1}) \bigr )}
 {\displaystyle \prod_{\overset{r=1}{r \neq i}}^{k}}\frac{(y_{n+1} - (1-\omega + n_r-\lambda_{_{(r)}}))}
{\bigl( (n_{i}-\lambda_{_{(i)}}) - (n_{r} - \lambda_{_{(r)}}) \bigr )} \, .
\end{array}
\ee
Using these formulas, as well as the initial data: $E_{T_{\{1; 1\}}} = 1$,  step by step we get explicit expressions
for all primitive idempotents corresponding to paths in the oscillating Young graph for the Brauer algebra $\mathcal{B}r_j(\omega)$.

{ \bf Example.} For the algebra $\mathcal{B}r_j$ we construct
 an explicit expression for the complete symmetrizer $E_{T_{\{[j];j\}}}$.
In the space of the left regular representation of the algebra $\mathcal{B}r_j$ the idempotent $E_{T_{\{[j-1];j-1\}}}$
extracts the subspace in which the element $y_j$ can be equal to
 three eigenvalues.
Indeed, according to the {\it branching rule} discussed above,
there are two possibilities to add a box to the
  diagram $[j-1]$, as indicated by the asterisks in the figure below, or delete the last box in the row with the content $(j-2):$
\begin{center}
\begin{tabular}{lllllll}
\cline{1-6}
\multicolumn{1}{|l|}{0} & \multicolumn{1}{l|}{1} & \multicolumn{2}{l|}{$\dots$} & \multicolumn{1}{l|}{$j-3$} & \multicolumn{1}{l|}{$j-2$} & $*$ \\ \cline{1-6}
$*$                       &                        &             &             &                          &                          &
\end{tabular}
\end{center}
Following the procedure for constructing idempotents described above,
we deduce the identity for the element $y_j$
\be \lb{haS}
E_{T_{\{[j-1];j-1\}}} \cdot (y_j - (j-1) ) \cdot (y_j+1)\cdot(y_j+\omega + j - 3) = 0
\ee
Since the symmetrizer $E_{T_{\{[j];j\}}}\,$ corresponds to the
 Young diagram obtained by adding the box with the content
 $(j-1)$, expression (\ref{haS}) gives us the recurrence relation for the
 symmetrizer $(j \geq 2)$
\be \lb{rS}
E_{T_{\{[j];j\}}} = E_{T_{\{[j-1];j-1\}}} \cdot \frac{(y_j+1)\cdot (y_j+\omega +n - 3)}{j\cdot(2j-4+\omega)}
\ee
Solving this recurrence relation, we obtain an explicit formula for the
 complete symmetrizer
\be \lb{rS1}
E_{T_{\{[j];j\}}} = \frac{(y_2 +1)\cdots(y_j+1)}{j!} \cdot \frac{(y_2+\omega-1)\cdot(y_3 +\omega) \cdots (y_j+\omega+j-3)}{\omega \cdot (2+\omega) \cdots (2j-4+\omega)} \; .
\ee
 One can check that (\ref{rS1}) obeys the identities ($\forall r =1,...,j-1$)
 \be \lb{rS2}
\sigma_r \cdot E_{T_{\{[j];j\}}} =  E_{T_{\{[j];j\}}} \cdot \sigma_r
 = E_{T_{\{[j];j\}}}\; , \;\;\;\;
 \kappa_r \cdot E_{T_{\{[j];j\}}} = 0 =  E_{T_{\{[j];j\}}} \cdot \kappa_r  \; .
\ee
 The proof of these identities for the general
 $q$-deformed case is given in \cite{Isa1}.

\section{Realization of the Brauer algebra in tensor spaces}
\setcounter{equation}{0}

In the previous section, we used the presentation
of the Brauer algebra which is given in Definition 2. Note that
the Brauer algebra has a convenient visualisation
 in terms of braids (see \cite {BR}, \cite {IM}).
This graphical visualisation comes from a natural
representation (see e.g. \cite {IR}, \cite {BR}) in which
 generators (\ref{drel}) of the algebra $\mathcal{B}r_j$ 
 are given by some operators in the tensor space 
  $(\mathbb{R}^{p,D-p})^{\otimes j}$, where $\mathbb{R}^{p,D-p}$ is 
 the space of the defining representation of the orthogonal 
  group $SO(p,D-p)$.
In this section, we construct a family of tensor
representations of the Brauer algebra in the space 
 $(\mathbb{R}^{p,D-p})^{\otimes j}$.
The natural tensor
representation of the Brauer algebra is an
 example in this family. Another special
 representation in this family is used below
for constructing TT-operators.

Introduce the triple $(\theta,\hat{\theta},\Check{\theta})$
 of $D \times D$  real matrices
 $\Hat{\theta}= ||\Hat{\theta}_{n m}||$, $\Check{\theta} = ||\Check{\theta}^{n m}||$ and
$\theta = ||\theta^{n}_{\;\; m}|| = ||\theta^{\;\; n}_{m}||$
such that
 \be
 \lb{theta3}
 \Check{\theta}^{nm} \, \Hat{\theta}_{m\ell} = \theta^{n}_{\;\; \ell} \; , \;\;\;
 \Hat{\theta}_{\ell m} \, \Check{\theta}^{mn} = \theta^{\;\; n}_\ell \; , \;\;\;
 \Check{\theta}^{m\ell} \cdot \theta^{\;\; n}_{\ell}  = \Check{\theta}^{mn}  \; , \;\;\;
 \Hat{\theta}_{m\ell} \cdot \theta^\ell_{\;\; n}  = \Hat{\theta}_{mn} \; .
 \ee
 \begin{proposition}
The triple $(\theta, \, \Hat{\theta}, \, \Check{\theta})$
  with relations (\ref{theta3}) satisfy the conditions
\be
 \lb{theta1}
\theta^{\;\; m}_{\ell} \Hat{\theta}_{m n} = \Hat{\theta}_{\ell n} \; , \;\;\;
\theta^{\ell}_{\;\; m} \, \Check{\theta}^{m n}   = \Check{\theta}^{\ell n} \; , \;\;\;
 \theta^{n}_{\;\; r} \theta^{\,r}_{\;\; \ell} = \theta^{n}_{\;\; \ell} \; ,
 \ee
 \be
 \lb{theta2}
 {\rm Tr}(\theta) = \theta^{\;\; \ell}_\ell =
 \Hat{\theta}_{\ell m} \, \Check{\theta}^{m\ell} = \omega \; ,
 \ee
 where $\omega$ is an integer number: $0 \leq \omega \leq D$.
\end{proposition}
{\bf Proof.} Relations (\ref{theta1}) follow directly from equalities
(\ref{theta3}). Formula (\ref{theta2}) follows from the last relation
in (\ref{theta1}) which means that $\theta$ is a projector
 and eigenvalues of $\theta$ are equal to $0$ or $+1$. Thus, $\omega$
 is a rank of $\theta$. \hfill \qed

\vspace{0.2cm}

  Consider the operators $P_r^{(\theta)}$ and $K_r^{(\theta)}$ $(r=1,\dots,j-1)$ acting in the space
 $(\mathbb{R}^D)^{\otimes j}$ according to the following formulas:
\be \label{defPK}
\begin{array}{c}
P_r^{(\theta)} \cdot (e_{i_1} \otimes \cdots \otimes e_{i_r} \otimes e_{i_{r+1}} \otimes \cdots \otimes e_{i_j}) = \\[0.5cm]
(e_{\ell_1} \otimes \cdots \otimes e_{\ell_r} \otimes e_{\ell_{r+1}} \otimes \cdots \otimes e_{\ell_j})\theta^{\ell_1}_{i_1}
\cdots   \theta^{\ell_{r-1}}_{i_{r-1}}  \theta^{\ell_r}_{i_{r+1}}  \theta^{\ell_{r+1}}_{i_r}  \theta^{\ell_{r+2}}_{i_{r+2}} \cdots  \theta^{\ell_j}_{i_j} \; , \\[0.5cm]
K_r^{(\theta)} \cdot (e_{i_1} \otimes \cdots \otimes e_{i_r} \otimes e_{i_{r+1}} \otimes \cdots \otimes e_{i_j}) = \\[0.5cm]
(e_{\ell_1} \otimes \cdots \otimes e_{\ell_r} \otimes e_{\ell_{r+1}} \otimes \cdots \otimes e_{\ell_j})\theta^{\ell_1}_{i_1}
\cdots  \theta^{\ell_{r-1}}_{i_{r-1}} \Check{\theta}^{\, \ell_r \ell_{r+1}}
\Hat{\theta}_{i_r i_{r+1}} \theta^{\ell_{r+2}}_{i_{r+2}} \cdots \theta^{\ell_j}_{i_j} \; ,
\end{array}
\ee
where $i_\ell =1,...,D$ and $e_i$  are the basis vectors in the space $\mathbb{R}^D$.

\begin{proposition}\label{proS1}
If the triple $(\theta,\hat{\theta},\Check{\theta})$ satisfies
(\ref{theta3}), (\ref{theta1}) and (\ref{theta2}), then
the map $S_\theta$:
 ${\cal B}r_j(\omega) \to {\rm End}(\mathbb{R}^D)^{\otimes j}$ defined
 on the generators $\sigma_r,\kappa_r \in {\cal B}r_j(\omega)$:
\be \label{nrepBr2}
\begin{array}{c}
S_\theta (\sigma_r) = P_r^{(\theta)} \; , \;\;\;\;
S_\theta (\kappa_r) = K_r^{(\theta)} \; ,
\end{array}
\ee
is extended to the whole algebra ${\cal B}r_j(\omega)$ as a homomorphism
(i.e. $S_\theta$ is a representation of ${\cal B}r_j(\omega)$).
\end{proposition}
{\bf Proof.} Making use of formulas (\ref{theta3}), (\ref{theta1}) and (\ref{theta2}),
  one can check directly that the operators
 (\ref{nrepBr2}) satisfy the defining relations (\ref{drel}). It
 means that the map (\ref{nrepBr2}) can be extended
 to the whole algebra ${\cal B}r_j(\omega)$ as a homomorphism
 and $S_\theta$ is a representation of ${\cal B}r_j(\omega)$.  \hfill \qed

\vspace{0.2cm}

\noindent
{\bf Remark 1.} Consider the triple of $D \times D$ matrices
 $\theta^r_n = \delta^r_n$, $\hat{\theta}_{rn} = \eta_{rn}$,
 $\Check{\theta}^{rn} = \eta^{rn}$, where $\delta^{\ell}_i$ is
 the Kronecker delta and 
 \be \label{metret}
 ||\eta_{rn}|| = {\rm diag} 
 (\underbrace{1,...,1}_p,\underbrace{-1,...-1}_q) \;\;\;\;\;\;\;
 (p+q=D)
 \ee
  is the metric in $\mathbb{R}^{p,q}$. This triple
  satisfies (\ref{theta3}), (\ref{theta1}) and (\ref{theta2})
  with $\omega = D$. Thus, according to Proposition {\bf \ref{proS1}},
  this triple defines the
 representation $S$ of the Brauer algebra $\mathcal{B}r_j(D)$,
which acts in the space $(\mathbb{R}^{p, q})^{\otimes j}$
 (see (\ref{defPK}), (\ref{nrepBr2}))
\be \label{repBr1}
\begin{array}{c}
S(\sigma_r) \cdot (e_{i_1} \otimes \cdots \otimes e_{i_r} \otimes e_{i_{r+1}} \otimes \cdots \otimes e_{i_j}) = \\[0.5cm]
(e_{\ell_1} \otimes \cdots \otimes e_{\ell_r} \otimes e_{\ell_{r+1}} \otimes \cdots \otimes e_{\ell_j})\delta^{\ell_1}_{i_1}
\cdots   \delta^{\ell_{r-1}}_{i_{r-1}} \;
 \delta^{\ell_r}_{i_{r+1}}  \delta^{\ell_{r+1}}_{i_r}  \;
 \delta^{\ell_{r+2}}_{i_{r+2}} \cdots  \delta^{\ell_j}_{i_j} \; , \\[0.5cm]
S(\kappa_r) \cdot (e_{i_1} \otimes \cdots \otimes e_{i_r} \otimes e_{i_{r+1}} \otimes \cdots \otimes e_{i_j}) = \\[0.5cm]
(e_{\ell_1} \otimes \cdots \otimes e_{\ell_r} \otimes e_{\ell_{r+1}} \otimes \cdots \otimes e_{\ell_j})\delta^{\ell_1}_{i_1}
\cdots  \delta^{\ell_{r-1}}_{i_{r-1}} \; \eta^{\ell_r \ell_{r+1}} \eta_{i_r i_{r+1}} \; \delta^{\ell_{r+2}}_{i_{r+2}} \cdots \delta^{\ell_j}_{i_j},
\end{array}
\ee
where $e_\ell$ are the basis vectors in $\mathbb{R}^{p,q}$.
It is known (see e.g. \cite{IR})
 that the action  of the algebra $\mathcal{B}r_j(D)$
 in the space of representation (\ref{repBr1})
 centralizes the action of the group $SO(p,q)$ in the same space $(R^{p,q})^{\otimes j}$
of the representation $T^{(j)} \equiv T^{\otimes j}$,
 where $T$ are the defining representations of $SO(p,q)$. It means that any operator
 $X$, which is a linear combination
\begin{equation} \lb{sio2}
X = \sum_{i=1}^{(2j-1)!!} x_i \, S(a_i),
\end{equation}
where summation runs over all basis elements $a_i \in \mathcal{B}r_j(D)$
and $x_i$ are complex coefficients, commutes with any
element $g \in SO(p,q)$ in the representation $T^{(j)}$:
\begin{equation} \lb{sio1}
X \cdot \bigl ( T^{(j)} (g) \bigr ) = \bigl ( T^{(j)} (g) \bigr ) \cdot X \, .
\end{equation}
In view of Schur's Lemma, it means that
the tensor product $T^{\otimes j}$ of the defining representations $T$
 of $SO(p,q)$ is reducible and
it can be decomposed into a direct sum of its 
 irreducible components.
Subspaces of these irreducible components are extracted from the space
 of the representation $T^{\otimes j}$ by acting of special
projectors, which are images (in the representation $S$
given by (\ref{repBr1})) of
 the primitive orthogonal idempotents of the Brauer algebra $\mathcal{B}r_j(D)$
 (for details see \cite{IR}).

 \vspace{0.2cm}

\noindent
{\bf Remark 2.} Let the triple of matrices
$(\theta, \, \Hat{\theta} \, , \Check{\theta})$ be
 \be
 \lb{examp2}
\theta^{\, m}_{n} = \Theta^{m}_{n}(k) \; , \;\;\;
\Check{\theta}^{\, n m} = \Theta^{n}_{\ell}(k) \, \eta^{\ell m} \; , \;\;\;
 \Hat{\theta}_{ m n} = \Theta^{\ell}_{m}(k) \, \eta_{\ell n} \; ,
 \ee
where the metric $\eta$ is
given in (\ref{metret}), the matrix $\Theta^{m}_{n}(k)$
is defined in (\ref{genT}) and
depends on the momentum $k \equiv \vec{k} \in \mathbb{R}^{p,q}$.
The triple (\ref{examp2}) satisfies (\ref{theta3}), (\ref{theta1}) and (\ref{theta2})
  with $\omega = (D-1)$. Thus, according
  to Proposition {\bf \ref{proS1}}, the operators (\ref{defPK}) with the choice of
(\ref{examp2}) define the representation $S_{(\vec{k})} \equiv S_{\Theta(k)}$
of $\mathcal{B}r_j(D-1)$. Just this representation of the Brauer
algebra is needed for us to construct TT-projectors.

Consider the image of the complete symmetrizer (\ref{rS1}) in the
representation $S_{(\vec{k})}$
 \be
 \lb{exa2a}
 \Theta_{{\{[j] ; j \}}} \equiv S_{(\vec{k})}(E_{T_{\{[j];j\}}} ) \; .
 \ee
The operator $\Theta_{{\{[j] ; j \}}}$ acts in the space $(\mathbb{R}^{p,q})^{\otimes j}$
 and, in view of (\ref{rS2}), satisfies the
  conditions
\be \label{trsy}
 \begin{array}{c}
\Theta_{{\{[j] ; j \}}} \cdot S_{(\vec{k})}(\sigma_{i, \ell}) =
 S_{(\vec{k})}(\sigma_{i, \ell}) \cdot \Theta_{{\{[j] ; j \}}} = \Theta_{{\{[j] ; j \}}} \; ,
 \\ [0.2cm]
\;\;\; \Theta_{{\{[j] ; j \}}}\cdot S_{(\vec{k})}(\kappa_{i, \ell})
 = S_{(\vec{k})}(\kappa_{i, \ell}) \cdot \Theta_{\{[j] ; j \}} = 0 \, ,
\end{array}
\ee
where the elements $\sigma_{i, \ell}, \kappa_{i, \ell} \in \mathcal{B}r_j$ are defined in (\ref{Sh}).
\begin{proposition}\label{svop2}
 The operator $\Theta_{{\{[j] ; j \}}}$ given in (\ref{exa2a}) is
 equal to the $D$-dimensional spin projection operator $\Theta^{(j)}(k)$
 (see Definition {\bf 1} and Proposition {\bf \em \ref{svop1}})
 \be
 \lb{svop2a}
 \Theta_{{\{[j] ; j \}}} = \Theta^{(j)}(k) \; .
 \ee
\end{proposition}
{\bf Proof.} The component form of conditions (\ref{trsy}) is
 \be
 \lb{trsy01}
 (\Theta_{{\{[j] ; j \}}})^{n_1 ...........n_j}_{r_1...r'_i...r'_\ell...r_j}
  \Theta^{r'_i}_{r_\ell} \Theta^{r'_\ell}_{r_i}  =
  \Theta^{n_i}_{n'_\ell} \Theta^{n_\ell}_{n'_i}
  (\Theta_{{\{[j] ; j \}}})^{n_1 ...n'_i...n'_\ell...n_j}_{\;\; r_1............r_j}
  = (\Theta_{{\{[j] ; j \}}})^{n_1 ......n_j}_{ r_1.......r_j} \; ,
 \ee
 \be
 \lb{trsy02}
 (\Theta_{{\{[j] ; j \}}})^{n_1 ...........n_j}_{r_1...r'_i...r'_\ell...r_j}
 \, \Theta^{r'_i r'_\ell} \Theta_{r_ir_\ell}  =
 \Theta^{n_i n_\ell} \Theta_{n'_i n'_\ell} \,
 (\Theta_{\{[j] ; j \}})^{n_1 ...n'_i...n'_\ell...n_j}_{\;\; r_1............r_j} = 0
 \; , \;\;\;\;\; (\forall i,\ell) \, .
 \ee
 We contract relations (\ref{trsy01}) with
 momentum $k^{r_i}$ and $k_{n_i}$
 and use the properties $\Theta^{n}_{r}k^r = k_n \Theta^{n}_{r}=0$. As a result,
 we obtain  from relations (\ref{trsy01}) that the
 operator $\Theta_{{\{[j] ; j \}}}$ obeys
 properties 2.) and 3.) of Definition {\bf 1} in Section {\bf \ref{BF-proj}}.
 Then, we substitute explicit forms (\ref{genT}) of the matrices $\Theta(k)$
 in equations (\ref{trsy02}) and use the transversality property 3.) for the
 operator $\Theta_{{\{[j] ; j \}}}$. In this way, we deduce
  that $\Theta_{{\{[j] ; j \}}}$ obeys the
 property 4.) of Definition {\bf 1} in Section {\bf \ref{BF-proj}}. Since
 the symmetrizer (\ref{rS1}) is idempotent and satisfies the projection identity
 $E_{T_{\{[j];j\}}}^2 = E_{T_{\{[j];j\}}}$, the
 property 1.) of Definition {\bf 1} is fulfilled  automatically for the matrix
 $\Theta_{{\{[j] ; j \}}}$. Thus, the operator $\Theta_{{\{[j] ; j \}}}$
 obeys all four conditions of Definition {\bf 1} in Section {\bf \ref{BF-proj}},
 which determine the spin projection operator uniquely
 (see Proposition {\bf \em \ref{svop1}}).
 It leads to identity (\ref{svop2a}).  \hfill \qed

In this section, we have constructed the representations
$S_{(\vec{k})}$ of the Brauer algebra. These represen\-tations 
 act in spaces of the tensor representations
 of the orthogonal groups. The primitive ortho\-gonal idempotents
  of the Brauer algebra taken in these representations correspond
  to the spin projection operators, i.e. to the TT-projectors
  on the spaces of the irreducible representations of the $D$-dimen\-sional
  Poincar\'{e} group. The advantage of our approach compared to other approaches is that we automatically obtain TT-projectors
  with a given symmetry related to a certain Young diagram.
  Also, using the representation  (homomorphic map) $S_{(\vec{k})}$,
  one can transform the nontrivial identities for the
  Brauer algebra idempotents to the identities for
  the spin projection operators. For example, in the next section
we deduce, following this idea, a new factorized formula
for the Behrends-Fronsdal complete 
symmetrizer (\ref{genT02}), (\ref{genT11}).

\section{New factorization formula for Behrends-Fronsdal \\ symmetrizer.
 Recurrence relation}

Now we describe another approach  to
construct the completely symmetric
projectors $\Theta_{{\{[j] ; j \}}}$. This
 approach is based on the different construction (see \cite{IM})
  of the complete symmetrizer (\ref{rS1}) in the Brauer algebra $\mathcal{B}r_{j}$.

  Consider
the rational function $\hat{R}_i (w)$ with values in the algebra $\mathcal{B}r_{j}$
\be \lb{defRh}
\hat{R}_i (w) = \sigma_i(1-\frac{\sigma_i}{w} + \frac{\kappa_i}{w-\varkappa}), \;\;\; \varkappa = \frac{\omega}{2} - 1 \, ,
\ee
where the argument $w$ is usually called the {\bf spectral parameter}.
 This function is a solution of the Yang-Baxter equation
 in a braid group form
\be \lb{bybeq1}
\hat{R}_i (w)\hat{R}_{i+1} (w+v) \hat{R}_i (v) = \hat{R}_{i+1} (v) \hat{R}_i (w+v) \hat{R}_{i+1} (w)\, .
\ee
Note that for $w = -1$ the function $\hat{R}_i (-1)$ has the following properties:
 \be
 \lb{prsm1}
\sigma_i \hat{R}_i (-1) = \hat{R}_i (-1) \sigma_i = \hat{R}_i (-1), \;\;\; \kappa_i \hat{R}_i (-1) = \hat{R}_i (-1) \kappa_i = 0.
\ee
Define the element $\Xi_j \in \mathcal{B}r_{j}$ by means of the recurrence relations
 \be
 \lb{reqT1}
\Xi_j = \Xi_{j-1} \left( \prod_{i = j-1}^{1} \hat{R}_i (-i) \right) =
 \left( \prod_{i=1}^{j-1}\hat{R}_i(-i) \right) \Xi_{j-1} \, ,
\ee
where $\Xi_1 = 1$. Here and
below we use the following convention
 in the ordering of products of
 noncommutative operators:
 $$
 (\prod\limits_{i = j-1}^{1} \hat{R}_i) \equiv \hat{R}_{j-1}
 \cdots \hat{R}_{2} \hat{R}_{1} \;\; , \;\;\;\;\;\;\;\;
  (\prod\limits_{i = 1}^{j-1} \hat{R}_i) \equiv \hat{R}_{1}
 \hat{R}_{2} \cdots  \hat{R}_{j-1} \;\; .
 $$
\newline
Let us define the following elements:
\be \lb{tin1}
R_{i k} (w) = \sigma_{i, k} \cdot  \hat{R}_{i k}(w), \;\;\; R_{i k}(w)
 =  (1-\frac{\sigma_{i,k}}{w} +
 \frac{\kappa_{i, k}}{w-\varkappa}) \;\; \in \;\; {\cal B}r_j(\omega) \; ,
\ee
where $\sigma_{i, k}$ and $\kappa_{i, k}$ are defined in (\ref{Sh});
 the parameter $\varkappa$ is given in (\ref{defRh}).
 Then the following relation holds:
\be \lb{tin3}
\Xi_j = \prod^{1}_{k=j-1} \left ( \prod^{k}_{i=1} \sigma_i \right ) \cdot \prod^{2}_{i=j} \left ( \prod^{1}_{k=i-1} R_{k i} (k-i) \right ) \; ,
\ee
which is directly deduced from (\ref{reqT1}) and (\ref{tin1}).
We note that the elements $R_{i k}$ from (\ref{tin1}) are solutions
 of the standard Yang-Baxter equation (cf. \ref{bybeq1})
\be \lb{tin2}
R_{i k}(w) R_{i \ell} (w+v) R_{k \ell} (v) =
 R_{k \ell} (v) R_{i \ell} (w+v) R_{i k}(w) \; .
\ee
 The image of the solution (\ref{tin1}) in the representation (\ref{repBr1}) is
 known as Zamolodchikov's $R$-matrix \cite{Zam}.

{\bf Remark 3.} Note that we can omit the dependence on the spectral parameters
 $(k-i)$ in the $R$-matrices $R_{k i} (k-i)$ in equation (\ref{tin3}),
  since all spectral parameters are restored from
 the lower indices of the elements $R_{ki}$. Moreover, if we
 fix the spectral parameters in (\ref{tin2}) as $w=(i-k)$ and $v=(k-\ell)$,
 then the Yang-Baxter equation (\ref{tin2}) can be written in a concise form
  \be \lb{tin2b}
R_{i k} \, R_{i \ell} \, R_{k \ell}  =
 R_{k \ell} \, R_{i \ell} \, R_{i k}  \; .
\ee

 \begin{proposition}\label{proxi}
For the element $\Xi_j$ the following representation holds \cite{IM}:
\be \lb{ybf1}
\Xi_j = \prod_{k=1}^{j-1} \Bigl ( \prod_{\ell = j-1}^k \, \hat{R}_{\ell} (\ell - j) \Bigl ) \equiv
\prod_{\ell = j-1}^1 \hat{R}_{\ell} (\ell - j)  \prod_{\ell = j-1}^2 \hat{R}_{\ell} (\ell - j)  \cdots \prod_{\ell = j-1}^{j-1} \hat{R}_{\ell} (\ell - j) \; .
\ee
 The element $\Xi_j$ defined in (\ref{reqT1}) and (\ref{ybf1})
 satisfies the conditions (cf. (\ref{rS2}))
 \be \lb{rS3}
\sigma_r \cdot \Xi_j =  \Xi_j \cdot \sigma_r
 = \Xi_j \; , \;\;\;\; \kappa_r \cdot \Xi_j = 0 =  \Xi_j \cdot \kappa_r
 \;\;\;\; (r=1,...,j-1) \; ,
\ee
 and we have the identity
 \be \lb{tuq2}
E_{T_{\{[j] ; j \}}} = \frac{1}{j!} \, \Xi_{j}  \; ,
\ee
where the idempotent $E_{T_{\{[j] ; j \}}}$ is given in (\ref{rS1}).
\end{proposition}
{\bf Proof.}
 We prove identity (\ref{ybf1}) for
  the element $\Xi_j$ defined in (\ref{reqT1}) by induction.
 First, we rewrite the right-hand side of eq. (\ref{ybf1}) in terms of the elements
 (\ref{tin1}):
\be \lb{zz0}
 \prod_{k=1}^{j-1} \Bigl ( \prod_{\ell = j-1}^k \,
 \hat{R}_{\ell} (\ell - j) \Bigl ) =\prod^{j-1}_{k=1} \left ( \prod_{i=j-1}^{k} \sigma_i \right ) \cdot \prod_{1 \leq i < k \leq j}R_{i k} \, ,
\ee
where we use the concise notation $R_{i k} \equiv R_{i k}(i-k)$
and define the double product as
\be \lb{zz4}
\prod_{1 \leq i < k \leq j}R_{i k}  =
( \prod_{k=2}^j R_{1 k})\cdot (\prod_{k=3}^j R_{2 k}) \cdots
(\prod_{k=j-1}^{j} R_{j-2, k}) \cdot (\prod_{k=j}^{j} R_{j-1, k})\, .
\ee
 Then we note that the prefactors containing only elements $\sigma_i$
 in right-hand sides of (\ref{tin3}) and (\ref{zz0})
 are equal
\be \lb{zz2}
\prod^{1}_{k=j-1} \left ( \prod^{k}_{i=1} \sigma_i \right ) =
\prod^{j-1}_{k=1} \left ( \prod_{i=j-1}^{k} \sigma_i \right ) \; ,
\ee
  (one can prove this identity by means of induction over $j$).
 So to prove (\ref{ybf1}), it is sufficient to show that the
 following equality holds:
\be \lb{zz3}
\prod^{2}_{i=j} \left ( \prod^{1}_{k=i-1} R_{k i} \right )
= \prod_{1 \leq i < k \leq j}R_{i k} (i-k).
\ee
 For $j=3$ we have the base of induction
 $R_{23} R_{13} R_{12}= R_{12} R_{13} R_{23}$ which is
 valid in view of the Yang-Baxter equation (\ref{tin2b}).
 Below we need the identity
 \be \lb{zz7}
\left[ \prod^{k}_{i=j-1}  R_{ij} \right]
\left (  \prod^{j-1}_{\ell=k+1}  R_{k \ell} \right )
= \left (  \prod^{j}_{\ell=k+1}  R_{k \ell} \right )
\left[ \prod^{k+1}_{i=j-1}  R_{ij} \right] \; ,
\ee
 which can be deduced by applying the Yang-Baxter equation
 (\ref{tin2b}) many times.
 Suppose that (\ref{zz3}) is valid for $j \to (j-1)$. Then we
 consider the left-hand side of (\ref{zz3})
\be \lb{zz1}
\begin{array}{c}
\displaystyle \prod^2_{i=j} \left ( \prod^{1}_{k=i-1} R_{k i}\right) =
\prod^{1}_{i=j-1} R_{i j}  \cdot \prod_{_{1 \leq i < \ell \leq j-1}}R_{i \ell} =
 \\[0.7cm]
 \displaystyle
= \left[ \prod^{1}_{i=j-1} R_{i j} \right]
\cdot \Bigl(\prod^{j-1}_{\ell=2} R_{1, \ell}\Bigr) \cdot
\Bigl(\prod^{j-1}_{\ell=3} R_{2, \ell}\Bigr) \cdots
\Bigl(\prod^{j-1}_{\ell=j-2} R_{j-3, \ell}\Bigr) \cdot
\Bigl(\prod^{j-1}_{\ell=j-1} R_{j-2, \ell}\Bigr)  = \\[0.7cm]
\displaystyle = \Bigl(\prod^{j}_{\ell=2} R_{1, \ell}\Bigr) \cdot
\left[ \prod^{2}_{i=j-1} R_{i j} \right] \cdot
\Bigl(\prod^{j-1}_{\ell=3} R_{2, \ell}\Bigr)
\Bigl(\prod^{j-1}_{\ell=4} R_{3, \ell}\Bigr) \cdots
\Bigl(\prod^{j-1}_{\ell=j-1} R_{j-2, \ell}\Bigr)  =  \dots = \\[0.7cm]
\displaystyle \dots  = \Bigl(\prod^{j}_{\ell=2} R_{1, \ell}\Bigr) \cdot
\Bigl(\prod^{j}_{\ell=3} R_{2, \ell}\Bigr) \cdots
\Bigl(\prod^{j}_{\ell=j-1} R_{j-2, \ell}\Bigr) \cdot [ R_{j-1, j} ]
= \prod_{_{1 \leq i < \ell \leq j}}R_{i \ell}\, ,
\end{array}
\ee
where in the first equality we use the induction hypothesis and in the following
equalities we apply (\ref{zz7}) many times
 (everywhere we use the concise notation from Remark {\bf 3}).

 Now the conditions (\ref{rS3}) are deduced from the representation
  (\ref{ybf1}), recurrence relations (\ref{reqT1}) and
 conditions (\ref{prsm1}). Finally, we prove formula (\ref{tuq2}).
 Since the properties (\ref{rS2}) for the idempotent $E_{T_{\{[j] ; j \}}}$
 are the same as the properties (\ref{rS3}) for the element $\Xi_j$, then
 $\Xi_j$ has to be proportional to the idempotent $E_{T_{\{[j] ; j \}}}$.
 Therefore, to prove (\ref{tuq2}), it is enough to show that
 the element in the right-hand side of (\ref{tuq2}) satisfies
 the projection property, i.e. we need to verify the identity
  $\Xi_{j} \cdot \Xi_{j} = j! \; \Xi_j$. Indeed,
   it follows from the chain of equalities
\be \lb{pp1}
\begin{array}{c}
\Xi_{j} \cdot \Xi_{j}  = \Xi_{j-1} \cdot \hat{R}_{j-1} \cdots \hat{R}_{2} \cdot \hat{R}_{1} \cdot \Xi_{j}
= j \cdot  \Xi_{j-1} \cdot \Xi_{j} = \\[0.5cm]
= j \cdot  \Xi_{j-2} \cdot \hat{R}_{j-2} \cdots \hat{R}_{1} \cdot \Xi_{j}
= j (j-1) \cdot  \Xi_{j-3} \cdot \hat{R}_{j-3} \cdots \hat{R}_{1} \cdot \Xi_{j}
= \dots = j! \cdot \Xi_j \, ,
\end{array}
\ee
 where $\hat{R}_{k} \equiv \hat{R}_{k}(-k)$.
Here we use the recurrence
formula (\ref{reqT1}) for the elements $\Xi_\ell$ and then apply the relations
 $$
\hat{R}_{k}(-k) \, \Xi_{j}  = \frac{k+1}{k}  \, \Xi_{j} \; , \;\;\;\;\;
\forall k=1,..., (j-1) \; ,
 $$
 where we take into account the explicit form (\ref{defRh})
 of the element $\hat{R}_k(-k)$ and conditions (\ref{rS3}).
\hfill \qed

 \vspace{0.2cm}

\noindent
 {\bf Remark 4.} Relations (\ref{pp1}) lead to the formula
 $(j-1)! \Xi_{j} = \Xi_{j-1} \cdot \Xi_{j}$. Making use of the
 second part of (\ref{reqT1}) in the right-hand side of this formula,
 we obtain a new recurrence identity for the symmetrizers
  $$
 \Xi_{j}  = \frac{1}{(j-2)!} \; \Xi_{j-1} \, \hat{R}_{j-1} \, \Xi_{j-1} \; ,
 $$
 which is useful in many applications.

\vspace{0.3cm}

  In view of identity (\ref{tuq2}) the element $(j!)^{-1} \, \Xi_j$ in
  the representation $S_{(\vec{k})}$ (see Remark {\bf 2}) is equal to the
  $D$-dimensional Behrends-Fronsdal symmetrizer (\ref{exa2a})
\be \lb{tuq1}
 \frac{1}{j!} S_{(\vec{k})}(\Xi_{j}) = \Theta^{(j)} \equiv
 \Theta_{{\{[j] ; j \}}}  \;
\ee
 where the generating function for the matrix $\Theta^{(j)}$ is given in
  (\ref{genT02}).

Now we prove identities (\ref{tuq1}) directly without using properties
 of Definition {\bf 1} in Sect. {\bf \ref{BF-proj}}.  For this purpose,
 we show that the matrices $(j!)^{-1} S_{(\vec{k})}(\Xi_{j})$ and $\Theta^{(j)}$
 satisfy the same recurrence relations.

 These recurrence relations have a unique solution and this
 means that equality (\ref{tuq1}) holds.

Introduce the generating function for the matrix $(j!)^{-1}S_{(\vec{k})}(\Xi_j)$
(cf. (\ref{genT01}))
 \be
 \lb{genT05}
\Xi^{(j)}(x,u) = \frac{1}{j!} \, u_{n_1} \cdots u_{n_j} \,
 (S_{(\vec{k})}(\Xi_j))^{n_1 \dots n_j}_{r_1 \dots r_j} \,
 x^{r_1} \cdots x^{r_j} \; ,
 \ee
 where the representation $S_{(\vec{k})}$ is defined in (\ref{defPK}), (\ref{examp2}).
\begin{proposition} \lb{CH1}
 For the generation function (\ref{genT05}) the following recurrence
 relation holds (cf. (\ref{shreq4})):

\be \lb{shreq5}
\begin{array}{c}
\Xi^{(j)}(x,u) = \frac{1}{(j-1)!}
\Bigl( \Theta^{(x)}_{(u)}
- \,\, \frac{1}{(\omega + 2 (j-2))} \, \Theta^{(x)}_{(x)} \,
 (u_k \, \partial_{x_k}) \Bigr)
\bigl ( \Theta^{(\,x\,)}_{(\partial_z)} \bigr)^{j-1}
\, \Xi^{(j-1)} (z,u)\,,
\end{array}
\ee

where  $\partial_{x_k} = \frac{\partial}{\partial x_k}$,
the function $\Theta^{(x)}_{(u)}$ is defined in (\ref{genT}) and $\omega = (D-1)$.
\end{proposition}
{\bf Proof.} Formula (\ref{shreq5}) is a consequence of the last equality in (\ref{reqT1}).
 Introduce the notation
 $$
 \tau_i := \sigma_i - \frac{2}{(\omega + 2(i - 1) )}  \kappa_{i}
 \;\;\;\;\;\; \Rightarrow \;\;\;\;\;\;
 \hat{R}_i(-i) = \left(  \frac{1}{i} + \tau_i \right) \; .
 $$
Then one can write the last equality in (\ref{reqT1}) in the form
\be \lb{newsh3}
 \Xi_j = {\amalg \!\! \amalg}_{j-1} \, \Xi_{j-1} \; ,
 \ee
where we have taken into account the conditions (\ref{rS3}) and introduce a
1-shuffle element (about shuffle elements
see \cite{IsOg09} and \cite{Isa1} and references therein)
\be \lb{newsh2}
 \begin{array}{c}
{\amalg \!\! \amalg}_{j-1}  =
\bigl ({1}+{\tau}_{j-1} + \tau_{j-2} \tau_{j-1} + \dots + \tau_{j-k} \cdots \
\tau_{j-2} \tau_{j-1} + \dots
+\tau_1 \tau_2 \cdots \tau_{j-1} \bigr)  \; .
 \end{array}
 \ee
Now in the representation $S_{(\vec{k})}$, identity (\ref{newsh3}) is written
for the generating function (\ref{genT05}) as
\be \lb{shreq3}
\begin{array}{c}
\Xi^{(j)}(x,u) = \frac{1}{j!} x^{n_1} \cdots x^{n_j} \,
(\overline{\amalg \!\! \amalg}_{j-1})^{\ell_1 \dots \ell_{j-1} \ell_j}_{n_1 \dots n_{j-1} n_j} \,
\frac{\partial}{\partial z^{\ell_1}} \cdots \frac {\partial} {\partial z^{\ell_{j-1}}} \, \frac{\partial}{\partial t^{\ell_{j}} }\, \Xi^{(j-1)}(z,u) \, \Theta(t, u).
\end{array}
\ee
Here we define the element
\be \lb{newsh}
\overline{\amalg \!\! \amalg}_{j-1}  =
\bigl (\tilde{1}+\tilde{\tau}_{j-1} + \tilde{\tau}_{j-2} \tilde{\tau}_{j-1} + \dots + \tilde{\tau}_{j-k} \cdots \tilde{\tau}_{j-2} \tilde{\tau}_{j-1} + \dots
+\tilde{\tau}_1 \tilde{\tau}_2 \cdots \tilde{\tau}_{j-1} \bigr) \, .
\ee
where $\tilde{1} = \Theta_{n_1}^{l_1} \cdots \Theta_{n_j}^{l_j} $ and the element $\tilde{\tau}_{i}$ has the form
\footnote{Here the operator $\tilde{1}$ really plays the role of a unit, since it trivially
acts on all covariant combinations constructed from the matrix $\Theta=||\Theta^{\,n}_{m}||$ and space-time metric $\eta = ||\eta_{nm}||$ .}
\be \lb{dtau1}
\begin{array}{c}
(\, \tilde{\tau}_i \,)_{n_1 \dots n_j}^{l_1 \dots l_j} = \Bigl ( \Theta_{n_1}^{l_1} \cdots  \Theta_{n_{i+1}}^{l_i}  \Theta_{n_i}^{l_{i+1}} \cdots \Theta_{n_j}^{l_j}
- \frac{2}{ (\omega + 2(i - 1) )} \cdot
\Theta_{n_1}^{l_1} \cdots  \Theta_{n_i n_{i+1}}  \Theta^{l_i l_{i+1}} \cdots \Theta_{n_j}^{l_j}
\Bigr ),
\end{array}
\ee
where  $\omega = D-1$.
Let us simplify formula (\ref{shreq3}).
Note, we will need it later, that the following formula hold:
\be \lb{rqsh}
\overline{\amalg \!\! \amalg}_{j-1}  =
\bigl(\tilde{1}+\overline{\amalg \!\! \amalg}_{j-2} \, \tilde{\tau}_{j-1} \bigr) \, .
\ee
Consider the differential operator
\be \lb{dish}
x^{n_1} \cdots x^{n_j} \,
(\overline{\amalg \!\! \amalg}_{j-1})^{\ell_1 \dots \ell_{j-1} \ell_j}_{n_1 \dots n_{j-1} n_j} \,
\frac{\partial}{\partial z^{\ell_1}} \cdots \frac {\partial} {\partial z^{\ell_{j-1}}} \, \frac{\partial}{\partial t^{\ell_{j}} }
\ee
from the right hand-side of (\ref{shreq3}).
Next we show
that (\ref{dish}) reduces to the following sum of two simple terms
\be \lb{dish1}
\begin{array}{c}
j \cdot \Theta^{(\,x\,)}_{(\partial_t)} \, \bigr ( \Theta^{(\,x\,)}_{(\partial_z)} \bigl )^{j-1} - \,\, \frac{j(j-1)}{(\omega + 2 (j-2))} \, \Theta^{(x)}_{(x)} \,
\Theta^{(\partial_t)}_{(\partial_z)} \, \bigl ( \Theta^{(\,x\,)}_{(\partial_z)} \bigr)^{j-2}
\end{array}
\ee
We now show that formula (\ref{dish1}) is equivalent to (\ref{dish}).
We carry out the proof by induction on $j$. For $j=2$ formula (\ref{newsh}) is represented as
\be \lb{in1}
(\overline{\amalg \!\! \amalg}_{1})^{r_1 r_2}_{n_1 n_2} = \Theta_{n_1}^{r_1} \Theta_{n_2}^{r_2} +  \Theta_{n_2}^{r_1} \Theta_{n_1}^{r_2}
- \frac{2}{\omega}\, \Theta_{n_1 n_2} \Theta^{r_1 r_2}
\ee
We make a contraction similar to (\ref{dish}) for $j=2$, using formula (\ref{in1})
and the definition of the generating function $\Theta^{(x)}_{(u)}$, as a result we have
\be \lb{in2}
x^{n_1} x^{n_2} (\overline{\amalg \!\! \amalg}_{1})^{r_1 r_2}_{n_1 n_2} \frac{\partial}{\partial z^{r_1}} \frac{\partial}{\partial t^{r_2}} =
2 \cdot \Theta^{(\,x\,)}_{(\partial_t)} \, \Theta^{(\,x\,)}_{(\partial_z)}-  \frac{2}{\omega} \, \Theta^{(x)}_{(x)} \, \Theta^{(\partial_t)}_{(\partial_z)} ,
\ee
it follows that formulas  (\ref{dish1}) and (\ref{dish}) are equivalent for $j=2$.
 Now we consider the sequence of transformations for (\ref{dish})
in case any $j$
\be \lb{in3}
\begin{array}{c}
x^{n_1} \cdots x^{n_j} \,
(\overline{\amalg \!\! \amalg}_{j-1})^{\ell_1 \dots \ell_{j-1} \ell_j}_{n_1 \dots n_{j-1} n_j} \,
\frac{\partial}{\partial z^{\ell_1}} \cdots \frac {\partial} {\partial z^{\ell_{j-1}}} \, \frac{\partial}{\partial t^{\ell_{j}} } =
x^{n_1} \cdots x^{n_j} \, \bigl ( \Theta_{n_1}^{\ell_1} \cdots \Theta_{n_j}^{\ell_j} + \\ [0.5cm]
(\overline{\amalg \!\! \amalg}_{j-2})^{\ell_1 \dots \ell_{j-2} r_{j-1}}_{n_1 \dots n_{j-2} n_{j-1}} \Theta_{n_j}^{r_j} (
  \Theta_{r_j}^{\ell_{j-1}} \Theta_{r_{j-1}}^{\ell_j}
- \frac{2}{(\omega +2(j-2) )}\,  \Theta_{r_{j-1} r_j} \Theta^{\ell_{j-1} \ell_j} ) \bigr)
\frac{\partial}{\partial z^{\ell_1}} \cdots \frac {\partial} {\partial z^{\ell_{j-1}}} \, \frac{\partial}{\partial t^{\ell_{j}} } = \\[0.5cm]
=\Theta^{(\,x\,)}_{(\partial_t)} \, \bigr ( \Theta^{(\,x\,)}_{(\partial_z)} \bigl )^{j-1}
\, + \, (j-1) \Bigl( \Theta^{(\,x\,)}_{(\partial_c)} \, \bigr ( \Theta^{(\,x\,)}_{(\partial_z)} \bigl )^{j-2} - \,\, \frac{(j-2)}{(\omega + 2 (j-3))} \, \Theta^{(x)}_{(x)} \,
\Theta^{(\partial_c)}_{(\partial_z)} \, \bigl ( \Theta^{(\,x\,)}_{(\partial_z)} \bigr)^{j-3} \Bigr )  \\[0.5cm]
\cdot \Bigl (
  \Theta^{(\,x\,)}_{(\partial_z)} \Theta_{(\,c\,)}^{(\partial_t)}
- \frac{2}{(\omega +2(j-2) )}\,  \Theta_{(c)}^{(x)} \Theta^{(\partial_t)}_{(\partial_z)} \Bigr) = \Theta^{(\,x\,)}_{(\partial_t)} \, \bigr ( \Theta^{(\,x\,)}_{(\partial_z)} \bigl )^{j-1}
+(j-1)  \Bigl (  \Theta^{(\,x\,)}_{(\partial_t)}\, \bigr ( \Theta^{(\,x\,)}_{(\partial_z)} \bigl )^{j-1} -  \\[0.5cm]
- \bigl ( \frac{2}{(\omega +2(j-2) )}\,
+ \,\frac{(j-2)}{(\omega + 2 (j-3))} - \frac{2 (j-2)} {(\omega +2(j-2) ) (\omega + 2 (j-3))} \bigr) \, \Theta^{(x)}_{(x)} \,
\Theta^{(\partial_t)}_{(\partial_z)} \, \bigl ( \Theta^{(\,x\,)}_{(\partial_z)} \bigr)^{j-2} \Bigr )  = \\[0.5cm]
= j \cdot \Theta^{(\,x\,)}_{(\partial_t)} \, \bigr ( \Theta^{(\,x\,)}_{(\partial_z)} \bigl )^{j-1} - \,\, \frac{j(j-1)}{(\omega + 2 (j-2))} \, \Theta^{(x)}_{(x)} \,
\Theta^{(\partial_t)}_{(\partial_z)} \, \bigl ( \Theta^{(\,x\,)}_{(\partial_z)} \bigr)^{j-2} \, .
\end{array}
\ee
Here in the first equality we used the recurrence relation (\ref{rqsh}),
in the second equality we used the induction hypothesis  (equivalence of formulas  (\ref{dish1}) and (\ref{dish}) for $\overline{\amalg \!\! \amalg}_{j-2}$),
 and in the third equation we applied differentiation with respect to the auxiliary $D$-vector $c$.
Based on (\ref{in3}), we can conclude that formulas (\ref{dish}) and (\ref{dish1}) are equivalent.

Next, substituting (\ref{dish1}) in (\ref{shreq3}) and then taking the derivative with respect to the variable $t$, we get
\be \lb{oj0}
\begin{array}{c}
\Xi^{(j)}(x,u) = \frac{1}{(j-1)!} \left ( \Theta^{(\,x\,)}_{(\,u\,)} \, \bigr ( \Theta^{(\,x\,)}_{(\partial_z)} \bigl )^{j-1} - \,\, \frac{(j-1)}{(\omega + 2 (j-2))} \, \Theta^{(x)}_{(x)} \,
\Theta^{(\,u\,)}_{(\partial_z)} \, \bigl ( \Theta^{(\,x\,)}_{(\partial_z)} \bigr)^{j-2} \right ) \, \Xi^{(j-1)}(z,u) \, .
\end{array}
\ee
Now taking the expression $\bigr(\Theta^{(\, x \,)}_{(\partial_z)} \bigl)^{j-1}$ out of the bracket of  the right-hand side of (\ref{oj0}), we reach formula (\ref{shreq5}).
\hfill \qed

\section{Explicit examples of spin TT-projectors related to 
special Young diagrams}

In this section we give as an example explicit construction
of two idempotents of the Brauer algebra ${\cal B}r_j$.
 These idempotents correspond,
in the representation $S_{(\vec{k})}$ (see Section {\bf 4}),
 to the $D$-dimensional projectors of the Behrends-Fronsdal
type with symmetries, which are related to the Young diagrams
$[1^j]$ and $[2,1]$ with a number of
 rows greater than one. For definiteness, we fix 
 in the definition (\ref{examp2}) of the representation $S_{(\vec{k})}$
 the metric $\eta = {\rm diag}(+1,-1,...,-1)$ corresponding to
 the case of the group $ISO(1,D-1)$.

\vspace{0.2cm}

\noindent
{\bf Example 1.}
At the end of Section {\bf \ref{baa}}, we gave
explicit formula (\ref{rS1}) for the symmetrizer $E_{T_{\{[j];j\}}}$.
Here we construct the idempotent
$E_{T_{\{[1^j];j\}}}$ which is an antisymmetrizer and
 corresponds to the Young diagram
$\lambda = [1^j]$ consisting of one column of height $j$.
According to the general formula (\ref{harT}), we have the identity
for the antisymmetrizer $E_{T_{\{[1^{j-1}];j-1\}}}$ 
\be \lb{ci0}
E_{T_{\{[1^{j-1}];j-1\}}} \cdot (y_j -1) (y_j +j-1) (y_j +\omega -j + 1) = 0\, ,
\ee
where $y_j$ is the Jucys-Murphy element.
Using this identity and formula (\ref{EA}), we obtain the
 recurrence relation for the antisymmetrizer $E_{T_{\{[1^j];j\}}}$
\be \lb{ci1}
E_{T_{\{[1^j];j\}}} = \frac{E_{T_{\{[1^{j-1}];j-1\}}}  \cdot (1-y_j) (y_j + \omega - j + 1)}{j (-2j+\omega+2)} \, .
\ee
Solution of the recurrence relation (\ref{ci1}) with the initial condition
$E_{T_{\{[1];1\}}} = 1$ is
\be \lb{ci2}
E_{T_{\{[1^j];j\}}} = \frac{(1\!-\!y_2)(1\!-\!y_3) \dots (1\!-\!y_j)\cdot(y_2\!+\!\omega\!-\!1)(y_3\!+\!\omega -2) \dots (y_j\!+\!\omega\!-\!j\!+\!1)}{j!(\omega-2)(\omega-4)\dots(\omega-2(j-1)\,)}\, .
\ee
Note that for the antisymmetrizer $E_{T_{\{[1^j];j\}}}\in \mathcal{B}r_j$,
one can write a simpler formula than (\ref{ci2}). To derive this 
 formula, we write the conditions for $E_{T_{\{[1^j];j\}}}$,
 which are analogs of (\ref{rS2}):
\be \lb{ci3}
\begin{array}{c}
\sigma_r \cdot E_{T_{\{[1^j];j\}}} = - E_{T_{\{[1^j];j\}}} = E_{T_{\{[1^j];j\}}} \cdot \sigma_r \, ,
\end{array}
\ee
\be \lb{ci3b}
\kappa_r \cdot E_{T_{\{[1^j];j\}}} = 0 = E_{T_{\{[1^j];j\}}} \cdot \kappa_r \, , \;\;\; \forall r =1,...,j-1 \, .
\ee
It is clear that relations (\ref{ci3b}) follow
from the conditions (\ref{ci3}).
Indeed, we have
\be \lb{ci4}
\kappa_r \cdot E_{T_{\{[1^j];j\}}} = \kappa_r \cdot \sigma_r \cdot E_{T_{\{[1^j];j\}}} = -  \kappa_r \cdot E_{T_{\{[1^j];j\}}} \, \;\;
\Rightarrow \;\; \kappa_r \cdot E_{T_{\{[1^j];j\}}}= 0\, ,
\ee
 where we used the relations $\kappa_r = \kappa_r \sigma_r$ and
 (\ref{ci3}).
 The second equality is proved analogously.
 Therefore, the conditions (\ref{ci3}) completely
 determine the antisymmetrizer $E_{T_{\{[1^j];j\}}}$
 in the Brauer algebra. Note that the same 
 conditions determine the antisymmetrizer in the group algebra of the permutation group $\mathbb{C}[S_j]$.
 It means that the expression for 
 $E_{T_{\{[1^j];j\}}} \in {\cal B}r_j$
 does not include the elements $\kappa_i \in {\cal B}r_j$
 and has the form which is specific for
 $\mathbb{C}[S_j]$ (see for example \cite{IR})
\be \lb{ci5}
E_{T_{\{[1^j];j\}}} = \frac{1}{j!} (1 - \sigma_{1,2}) \cdot (1 - \sigma_{1,3} - \sigma_{2,3} ) \cdots (1-\sigma_{1,j} - \dots - \sigma_{j-1, j}) \, .
\ee
We expand the brackets in the right-hand side of eq. (\ref{ci5}) and
take it in the representation $S_{(\vec{k})}$. As a result, we obtain
the explicit expression for the {\bf \em completely antisymmetric}
 $D$-dimensional spin projector
(antisymmetric analog of the Behrends-Fronsdal projector)
\be \lb{ci6}
\bigl ( S_{(\vec{k})} (E_{T_{\{[1^j];j\}}}) \bigr )^{d_1 \dots d_j}_{n_1 \dots n_j} = \frac{1}{j!} \sum_{\sigma \in S_j} (-1)^{p(\sigma)} \Theta^{d_1}_{n_{\sigma(1)}} \Theta^{d_2}_{n_{\sigma(2)}} \cdots \Theta^{d_j}_{n_{\sigma(j)}} \, ,
\ee
where the sum runs over all elements $\sigma$ of the permutation group $S_j$ and $p(\sigma)$ is the parity of $\sigma$.
Note that the antisymmetric projector (\ref{ci6})
is (by construction) the TT-projector and
is orthogonal to the completely symmetric
Behrends-Fronsdal projector (\ref{genT02}).

\vspace{0.2cm}

\noindent
{\bf Example 2.} Here we
 consider the Brauer algebra idempotents $E_{T_{\{[2,1];3\}}}$,
 which
correspond to the hook Young diagram $\lambda = [2,1]$. 
 The images of the idempotents $E_{T_{\{[2,1];3\}}}$
  in the representation $S_{(\vec{k})}$ yield 
  the corresponding TT-projectors. To construct the idempotents
  $E_{T_{\{[2,1];3\}}}$ explicitly,
 we consider two oscillating Young tableaux
  $\Lambda_1$ and $\Lambda_2$ (see Section {\bf \ref{baa}})
\be \lb{kim1}
\Lambda_1 = \{ \emptyset \overset{a_1 = 0}{\rightarrow} [1]  \overset{a_2=1}{\rightarrow} [2]  \overset{a_3=-1'}{\rightarrow} [2,1] \} \, , \;\;
\Lambda_2 = \{ \emptyset \overset{a_1 = 0}{\rightarrow} [1]  \overset{a_2=-1'}{\rightarrow} [1,1]  \overset{a_3=1}{\rightarrow} [2,1] \} \,  .
\ee
Both tableaux have the length $3$ and in both
cases the final Young diagram is a hook
\begin{center}
[2,1] \;\; = \;\; \begin{tabular}{|l|llll}
\cline{1-2}
0  & \multicolumn{1}{l|}{1} &  &  &  \\ \cline{1-2}
-1 &                        &  &  &  \\ \cline{1-1}
\end{tabular}
\!\!\!\!\!\!\!\!\! .
\end{center}
Let $e_s$ and $e_a$ be primitive idempotents corresponding to the diagrams
$[2]$ and $[1^2]$, respectively, which have the following explicit expressions:
 \be \lb{ki1}
 e_s = \frac{(1+y_2)(y_2+\omega-1)}{2 \omega}, \;\;\; e_a = \frac{(1-y_2)(y_2+\omega-1)}{2(\omega-2)} \, .
\ee
These expressions
 can be easily obtained from the general formula (\ref{EA}). By
using the {\bf \em branching rules}
 (see Fig. \ref{Fig1}) and oscillating Young tableaux
 $\Lambda_1$ and $\Lambda_2$, given in (\ref{kim1}),
we derive two identities (\ref{harT}) for the elements $e_s$, $e_a$ and $y_3$
\be \lb{ki0}
e_s \cdot (y_3-2)(y_3+1)(y_3+\omega ) = 0, \;\;\; e_a \cdot (y_3+2)(y_3-1)(y_3+\omega-2) = 0 \, .
\ee
 By means of
 these identities we construct (see formulas (\ref{EA}) and (\ref{ED}))
  two primitive idempotents
\be \lb{ki2}
e_{\Lambda_1} = \frac{e_s \cdot (2-y_3)(y_3+\omega)}{3(\omega-1)}\, , \;\;\;
e_{\Lambda_2} = \frac{e_a \cdot (2+y_3)(y_3+\omega - 2)}{3(\omega-1)} \, ,
\ee
for which (by construction) we have
$e_{\Lambda_\alpha}^2=e_{\Lambda_\alpha}$ and
 $e_{\Lambda_1} \cdot e_{\Lambda_2}=0$.
Opening the brackets in (\ref{ki2}), we obtain
\be \lb{pk0}
\begin{array}{c}
e_{\Lambda_1} = \frac{1}{6 \omega (\omega-1)} \bigl (2 \omega(\omega-1) + 2 \omega^{2} y_2 + (\omega (3-\omega) -2) y_{3} + \omega(2-\omega) y_2 y_3 +\\[0.5cm]
+ 2 \omega y_2^2 + (1-\omega) y_3^2+(2-\omega) y_2^2 y_3 - \omega y_2 y_3^2 - y_2^2 y_3^2 \bigr ) \, ,
\end{array}
\ee
\be \lb{pk1}
\begin{array}{c}
e_{\Lambda_2} = \frac{1}{6(\omega-1)(\omega-2)} \bigl (2 ( \omega(\omega-3) + 2)+2 (\omega(4-\omega)-4) y_2 + \omega (\omega-1) y_{3} + \omega(2-\omega) y_2 y_3 +  \\[0.5cm]
+ 2(2- \omega) y_2^2 +(\omega-1) y_3^2 - \omega y_2^2 y_3 + (2-\omega) y_2 y_3^2 - y_2^2 y_3^2  \bigr ).
\end{array}
\ee
Taking into account the
  identity $\kappa_1 y_3 = 0$ and definition (\ref{jme25})
  for $y_2$, one can rewrite
   relations (\ref{pk0}), (\ref{pk1}) in a concise form
\be \lb{pk2}
\begin{array}{c}
e_{\Lambda_1} = \frac{1}{6 (\omega-1)} \Bigl ((1+\sigma_1)(2 \omega
 -(\omega-2) y_3 - y_3^2) -4 \kappa_1 \Bigr ) \, ,
\end{array}
\ee
\be \lb{pk3}
\begin{array}{c}
e_{\Lambda_2} = \frac{1}{6 (\omega-1)} \Bigl ( (1-\sigma_1) (2 (\omega-2)
 +\omega y_3 +y_3^2)  \Bigr ).
\end{array}
\ee
Now we use the definition (\ref{jme25}) for $y_3$ and
express idempotents (\ref{pk2}) and (\ref{pk3})
in terms of the generators $\sigma_i, \kappa_i \in \mathcal{B}r_3$ $(i =1,2)$:
\be \lb{pk4}
\begin{array}{c}
e_{\Lambda_1} = \frac{1}{6} \Bigl (2 - (\sigma_1\sigma_2+\sigma_2\sigma_1) -\sigma_1 \sigma_2 \sigma_1+2\sigma_1-\sigma_2 +
\frac{1}{(\omega-1)}\bigl (2(\kappa_1 \kappa_2 + \kappa_2 \kappa_1) + \\[0.5cm]
+2 (\kappa_1 \sigma_2 +\sigma_2 \kappa_1)- (\kappa_2 \sigma_1+\sigma_1 \kappa_2 )
 -4 \kappa_1-\kappa_2  -\sigma_1 \kappa_2 \sigma_1 \bigr) \Bigr ) \, ,
\end{array}
\ee
\be \lb{pk5}
\begin{array}{c}
e_{\Lambda_2} = \frac{1}{6} \Bigl (2 -(\sigma_1\sigma_2+\sigma_2\sigma_1) +\sigma_1 \sigma_2 \sigma_1 - 2\sigma_1+\sigma_2
+\frac{1}{(\omega-1)}\bigl (3(\kappa_2 \sigma_1+\sigma_1 \kappa_2) - \\[0.5cm]
-3\kappa_2  -3 \sigma_1 \kappa_2 \sigma_1 \bigr ) \Bigr ).
\end{array}
\ee
 These are primitive idempotents in the algebra ${\cal B}r_3$,
 which are related to the Young diagram $[2,1]$.
The images of the idempotents
 $e_{\Lambda_1}$ and $e_{\Lambda_1}$ in the representation $S_{(\vec{k})}$
 have the form
\be \lb{pk6}
\begin{array}{c}
\bigl (S_{(\vec{k})} (e_{\Lambda_1}) \bigr)^{d_1 d_2 d_3}_{n_1 n_2 n_3}= \frac{1}{6} \Bigl ( 2 \, \Theta^{d_1}_{n_1} \Theta^{d_2}_{n_2}\Theta^{d_3}_{n_3}
- ( \Theta^{d_1}_{n_2} \Theta^{d_2}_{n_3} \Theta^{d_3}_{n_1} +  \Theta^{d_1}_{n_3} \Theta^{d_2}_{n_1} \Theta^{d_3}_{n_2} ) - \\[0.5cm]
- \Theta^{d_1}_{n_3} \Theta^{d_2}_{n_2} \Theta^{d_3}_{n_1} +  2\Theta^{d_1}_{n_2} \Theta^{d_2}_{n_1} \Theta^{d_3}_{n_3} -  \Theta^{d_1}_{n_1} \Theta^{d_2}_{n_3} \Theta^{d_3}_{n_2} - \\[0.5cm]
- \frac{1}{(\omega - 1)} \cdot \Bigl ( 4\Theta^{d_3}_{n_3} \Theta^{d_1 d_2}  \Theta_{n_1 n_2}
+\Theta^{d_1}_{n_1} \Theta^{d_2 d_3}  \Theta_{n_2 n_3}
+  \Theta^{d_2}_{n_2} \Theta^{d_1 d_3}  \Theta_{n_1 n_3}  - \\[0.5cm]
 -2( \Theta^{d_1}_{n_3} \Theta^{d_2 d_3}  \Theta_{n_1 n_2}
 +\Theta^{d_3}_{n_1} \Theta^{d_1 d_2}  \Theta_{n_2 n_3})
 -2(\Theta^{d_2}_{n_3} \Theta^{d_1 d_3}  \Theta_{n_1 n_2}
 +\Theta^{d_3}_{n_2} \Theta^{d_1 d_2}  \Theta_{n_1 n_3}) +  \\[0.5cm]
+ \Theta^{d_2}_{n_1} \Theta^{d_1 d_3}  \Theta_{n_2 n_3} + \Theta^{d_1}_{n_2} \Theta^{d_2 d_3} \Theta_{n_1 n_3}   \bigr )\Bigr ) \, ,
\end{array}
\ee
\be \lb{pk7}
\begin{array}{c}
\bigl (S_{(\vec{k})} (e_{\Lambda_2}) \bigr)^{d_1 d_2 d_3}_{n_1 n_2 n_3}= \frac{1}{6} \Bigl ( 2 \, \Theta^{d_1}_{n_1} \Theta^{d_2}_{n_2}\Theta^{d_3}_{n_3}
- ( \Theta^{d_1}_{n_2} \Theta^{d_2}_{n_3} \Theta^{d_3}_{n_1} +  \Theta^{d_1}_{n_3} \Theta^{d_2}_{n_1} \Theta^{d_3}_{n_2} ) + \\[0.5cm]
+ \Theta^{d_1}_{n_3} \Theta^{d_2}_{n_2} \Theta^{d_3}_{n_1} -  2\Theta^{d_1}_{n_2} \Theta^{d_2}_{n_1} \Theta^{d_3}_{n_3} +  \Theta^{d_1}_{n_1} \Theta^{d_2}_{n_3} \Theta^{d_3}_{n_2}- \\[0.5cm]
- \frac{3}{(\omega - 1)} \cdot \bigl ( \Theta^{d_1}_{n_1} \Theta^{d_2 d_3}  \Theta_{n_2 n_3}
+   \Theta^{d_2}_{n_2} \Theta^{d_1 d_3}  \Theta_{n_1 n_3}
-\Theta^{d_2}_{n_1} \Theta^{d_1 d_3}  \Theta_{n_2 n_3} - \Theta^{d_1}_{n_2} \Theta^{d_2 d_3} \Theta_{n_1 n_3}   \bigr )\Bigr ) \, .
\end{array}
\ee
It is clear that the operators
$S_{(\vec{k})} (e_{\Lambda_1})$ and $S_{(\vec{k})} (e_{\Lambda_2})$
given in (\ref{pk6}), (\ref{pk7}) are traceless and transverse
(see conditions 3 and 4 in Definition {\bf 1},
 Section {\bf \ref{BF-proj}}).
It means that
these operators are projectors onto two
irreducible representations of the group $ISO(1,D-1)$
acting in the space of 3-rank tensors.
These two representations are equivalent to each other.
 The explicit formulas for 3-rank tensors which form
 irreducible representations of the group $ISO(1,D-1)$
 are obtained by action of the TT-projectors 
 (\ref{pk6}) and (\ref{pk7}) on arbitrary tensors of rank 3.

 The sum of primitive idempotents (\ref{pk4}) and (\ref{pk5})
 gives the central idempotent 
 in the Brauer algebra $\mathcal{B}r_3$:
\be \lb{ki3}
 e_{_{[2,1]}} =  (e_{\Lambda_1} + e_{\Lambda_2}) \; ,
\ee
associated to the Young diagram $[2,1]$. The element $e_{_{[2,1]}}$
satisfies the relation $e_{_{[2,1]}}^2 = e_{_{[2,1]}}$ and
 belongs to the center
 because it is a symmetric function of the elements $y_2, y_3$ (it means
 that $e_{_{[2,1]}}$
commutes with all elements of the Brauer algebra $\mathcal{B}r_3$,
see \cite{IR}). To prove the last statement, we
substitute expressions (\ref{pk0}), (\ref{pk1}) of the idempotents
$e_{\Lambda_1}$ and $e_{\Lambda_2}$ into the right-hand side
of formula (\ref{ki3}).
As the result, we obtain the explicit form of $e_{_{[2,1]}}$ in terms
 of the Jucys-Murphy elements  $y_2$ and $y_3$ 
\be \lb{ki5}
\begin{array}{c}
e_{_{[2,1]}} =
\frac{1}{3} \Bigl ( 2-y_2 y_3 + \frac{1}{(\omega-1)} \bigl ( 2(y_2+y_3) - (y_2y_3^2 + y_2^2y_3\bigr) \Bigr ) \, ,
\end{array}
\ee
which is indeed a symmetric
 function of the elements $y_2$ and $y_3$.
In terms of the generators $\sigma_i, \kappa_i \in \mathcal{B}r_3$ $(i =1,2)$,
expression (\ref{ki5}) is written in the form
\be \lb{ki6}
\begin{array}{c}
e_{_{[2,1]}} = \frac{1}{3} \Bigl ( 2 -  \bigl (\sigma_1 \sigma_2 + \sigma_2 \sigma_1 \bigr ) +\frac{1}{(\omega-1)}
 \bigl (  (\kappa_1 \kappa_2 + \kappa_2 \kappa_1) \; + \\[0.5cm]
  + \; (\kappa_1 \sigma_2 + \sigma_2 \kappa_1)
  +  (\kappa_2 \sigma_1+ \sigma_1 \kappa_2)
- 2 \, (\kappa_1 +\kappa_2 + \sigma_1 \kappa_2 \sigma_1 ) \bigr) \Bigr )  \, .
\end{array}
\ee
Note that formula (\ref{ki6}) is mirror-symmetric,
 or in other words, it is invariant under simultaneous substitution:
$\sigma_1 \leftrightarrow \sigma_2$ and $\kappa_1 \leftrightarrow \kappa_2$. Finally, the image of the central idempotent
$e_{_{[2,1]}}$ in the representation $S_{(\vec{k})}$ has the form
\be \lb{ki8}
\begin{array}{c}
\bigl (S_{(\vec{k})} (e_{_{[2,1]}}) \bigr)^{d_1 d_2 d_3}_{n_1 n_2 n_3}= \frac{1}{3} \Bigl ( 2 \, \Theta^{d_1}_{n_1} \Theta^{d_2}_{n_2}\Theta^{d_3}_{n_3}
- ( \Theta^{d_1}_{n_2} \Theta^{d_2}_{n_3} \Theta^{d_3}_{n_1} +  \Theta^{d_1}_{n_3} \Theta^{d_2}_{n_1} \Theta^{d_3}_{n_2} ) - \\[0.5cm]
- \frac{1}{(\omega - 1)} \cdot \bigl ( 2 \cdot ( \Theta^{d_3}_{n_3} \Theta^{d_1 d_2}  \Theta_{n_1 n_2}
+ \Theta^{d_1}_{n_1} \Theta^{d_2 d_3}  \Theta_{n_2 n_3}
+  \Theta^{d_2}_{n_2} \Theta^{d_1 d_3}  \Theta_{n_1 n_3} ) - \\[0.5cm]
 - \Theta^{d_1}_{n_3} \Theta^{d_2 d_3}  \Theta_{n_1 n_2}
 -\Theta^{d_3}_{n_1} \Theta^{d_1 d_2}  \Theta_{n_2 n_3}
 -\Theta^{d_2}_{n_3} \Theta^{d_1 d_3}  \Theta_{n_1 n_2}
 -\Theta^{d_3}_{n_2} \Theta^{d_1 d_2}  \Theta_{n_1 n_3} -  \\[0.5cm]
-  \Theta^{d_2}_{n_1} \Theta^{d_1 d_3}  \Theta_{n_2 n_3} - \Theta^{d_1}_{n_2} \Theta^{d_2 d_3} \Theta_{n_1 n_3}   \bigr )\Bigr ) \, .
\end{array}
\ee
From the right-hand side of (\ref{ki8}) we see
 that $S_{(\vec{k})} (e_{_{[2,1]}})$
is transversal and traceless with respect to
 all pairs of lower and upper tensor indices
(Properties 3 and 4 in Definition {\bf 1},
 Section {\bf \ref{BF-proj}}), i.e. the
 following equalities hold:
\be \lb{ki7}
\begin{array}{c}
k^{n_1} \; \bigl (S_{(\vec{k})} (e_{_{[2,1]}})
\bigr)^{\;\; d_1 \; d_2 \; d_3}_{n_{_{\sigma(1)}} n_{_{\sigma(2)}} n_{_{\sigma(3)}}}
 = 0 =
k_{d_1} \; \bigl (S_{(\vec{k})} (e_{_{[2,1]}})
\bigr)^{d_{_{\sigma(1)}} d_{_{\sigma(2)}} d_{_{\sigma(3)}}}_{\;\;\; n_1 \; n_2 \; n_3} \, , \\ [0.4cm]
\eta^{n_1 n_2} \; \bigl (S_{(\vec{k})} (e_{_{[2,1]}})
\bigr)^{\;\; d_1 \; d_2 \; d_3}_{n_{_{\sigma(1)}} n_{_{\sigma(2)}} n_{_{\sigma(3)}}}
 = 0 =
\eta_{d_1 d_2} \; \bigl (S_{(\vec{k})} (e_{_{[2,1]}})
\bigr)^{d_{_{\sigma(1)}} d_{_{\sigma(2)}} d_{_{\sigma(3)}}}_{\;\;\; n_1 \; n_2 \; n_3} \, ,
\end{array}
\ee
where $\sigma$ is an arbitrary permutation from the group  $S_3$.
Thus, the operator (\ref{ki8}) is a TT-projector but its action
in the space of all 3-rank tensors gives a reducible representation 
of $ISO(1,D-1)$ since the projector (\ref{ki8}) 
in view of relation (\ref{ki3}) is the sum of two
primitive projectors (\ref{pk6}) and (\ref{pk7}).

\section{Conclusion.}
\setcounter{equation}0

In this paper, a new class of representations
 of the Brauer algebra is found.
This allows us to apply the method of constructing
 irreducible finite dimensional representations
of orthogonal and symplectic Lie groups (based on
using the idempotents of the Brauer algebra)
to construct irreducible representations
of the $D$-dimensional Poincar\'{e} group.
Using the new representations of the Brauer algebra,
we derive a new recurrence formula for 
$D$-dimensional completely symmetric BF projector.
In particular, we derive new explicit formulae
 for $D$-dimensional BF type projectors related to any
 symmetries, which correspond to the Young diagrams with two and
more rows (in contrast to fully symmetric BF projectors, which correspond to the single-row Young diagram). 
To illustrate the obtained results, we find images of
 some special idempotents of the Brauer algebra in the new representations.

We hope that the generalizations of the BF projectors obtained in this
paper will have useful
 applications. For example, generalized BF
projectors could be useful for constructing and investigating
different higher spin field theories.
In particularly, we know (see e.g. \cite{KP}), that the squares of BF spin
projectors are used as building blocks of invariant constructions
included in Lagrangians of higher spin field theories.

The formalism
described here for constructing irreducible representations 
of $ISO(1,D-1)$ with
a mixed type of symmetry is directly applied
to the case of (anti) de Sitter symmetry group. Namely, the
 Brauer algebra representation $S_{(\vec{k})}$ defined by triple
 (\ref{examp2}) can be formulated for metric $\eta$
 (\ref{metret})
 with any signature $(p,q)$ and in particular for 
 the case $p=2$ and $q=D-p$,
 which is specific for the group $ISO(2,D-2)$. We also note that
 this generalization seems possible, since some facts are already known in the literature
regarding the establishment of the correct correspondence between unitary 
irreducible representations of the groups $SO(2,D-2)$ and $SO(1,D-1)$
and fields, respectively, in the spaces $AdS_{D-1}$ and $dS_{D-1}$ 
(see papers \cite{RR}, \cite{JMP1}, \cite{JMP2},
\cite{BBB} and references therein).

\section*{Acknowledgments}

The authors would like to thank S.O.Krivonos and
O.V.Ogievetsky for useful discussions.
M.A.P. acknowledges the support of the Russian
Foundation for Basic Research, projects No.\,19-01-00726-a.
A.P.I. acknowledges the support of the Russian Science Foundation, grant No.\,19-11-00131.

 \vspace{1cm}

\begin{otherlanguage}{english}

\end{otherlanguage}

\end{document}